\documentstyle[aps,eqsecnum,epsfig]{revtex}
\tighten

\newcommand{\beq}[1]{
\begin{equation}\label{#1}}
\newcommand{\bea}[1]{
\begin{eqnarray}\label{#1}}

\begin{document}
\thispagestyle{empty}

\begin{flushright}
JLAB-THY-02-22 \\
hep-ph/0205315 \\
Revised \\
December 19, 2002
\end{flushright}

\vspace{1cm}
\begin{center}
{\Large \bf 
Power-Law  Wave Functions \\[2mm] and \\[3mm] 
Generalized Parton Distributions  for Pion}
\\[1cm]
\end{center}
\begin{center}
{ A. MUKHERJEE$^{a,b,c}$, I.V. MUSATOV$^{d}$, H. C. PAULI$^{e}$,
A. V. RADYUSHKIN$^{d,f}$\footnote{Also at Laboratory of Theoretical Physics,JINR, Dubna, Russia}
} \\[2mm]  
 {\em  $^a$Universit\"at Mainz, J. J. Becher Weg 45, D-55099, Mainz, Germany} \\[2mm]      
       {\em   $^b$Universit{\"{a}}t Regensburg, D-93040 Regensburg, Germany} \\[2mm]
       {\em  $^c$Saha Institute of Nuclear Physics, 1/AF Bidhan Nagar,
                Kolkata 700064, India} \\[2mm]
      {\em   $^d$ Theory Group, Jefferson Lab, Newport News, VA 23606, USA} \\[2mm] 
	{\em $^e$Max Planck Institut f{\"{u}}r Kernphysik, 69117 Heidelberg,
        Germany} \\[2mm]
       {\em  $^f$Physics Department, Old Dominion University, Norfolk, VA 23529,
USA.}
\end{center}
\vspace{1cm}

\begin{abstract}
We propose a model for generalized parton distributions of the 
pion based on the power-law {\it Ansatz} for the  effective 
light-cone wave function.  
  \\
  
  \vspace{.5cm}
\noindent {PACS number(s): 13.40.Gp, 13.60.Fz, 14.40.Aq}
\vspace{.5cm}
\end{abstract}
\vskip .2in
\newpage

\section{Introduction}
\vskip .2in

Generalized parton distributions (GPDs) 
\cite{Muller:1998fv,Ji:1996ek,Radyushkin:1996nd} 
are now 
the  object of intensive theoretical studies,
especially within the context of applications 
to deeply virtual \cite{Ji:1996nm,Radyushkin:1997ki} 
(for a
detailed recent review see Ref. \cite{Goeke:2001tz}) 
and large momentum transfer processes \cite{Radyushkin:1998rt,Diehl:1998kh}. 
The main advantage of GPDs is their universality
allowing  to connect different hard  processes,
both exclusive and inclusive. 
The price for this is the complexity of GPDs: they are functions
of 3 variables, {\it e.g.,} skewed parton distributions (SPDs) 
${\cal F}_\zeta (X,t)$ or $H(x,\xi;t)$
depend on the fraction $X$ (or $x$) 
of the momentum carried by the active 
quark, the skewedness parameter $\zeta$ (or $\xi$) 
and the invariant momentum transfer $t$.
For this reason, the most promising 
approach to  disentangling  GPDs from experimental data
is to construct realistic models 
for GPDs and fix their parameters by fitting the data.
The crucial point for the model building 
is that, 
in specific limits, GPDs reduce to more familiar
functions describing the hadronic structure,
such as usual parton densities, form factors 
and distribution amplitudes.
The ``reduction'' relations  between GPDs and 
these functions have been   used as a basis 
for building phenomenological models of GPDs \cite{Radyushkin:1998es}.
Another  fruitful idea used in the model-building is to construct 
GPDs from  the light-cone (LC) wave functions \cite{Radyushkin:1998rt,Diehl:1998kh}. 
The most popular   {\it {\it Ansatz}} \cite{huang} assumes a Gaussian dependence  
of the LC wave functions $\psi (x, k_{\perp})$ 
on the transverse momentum $k_{\perp}$.
A  pragmatic reason behind this choice is the simplicity
of Gaussian integrals allowing to obtain many results in  analytic form. 
However, there are no a priori grounds to exclude 
wave functions with other types of transverse  momentum 
dependence. In particular,  
the two-body ({\it i.e.,} $\bar qq$) component 
of the pion wave function was calculated recently in a model 
\cite{Frederico:2001qy} based on 
the one-gluon exchange approximation in the light-front framework. 
The wave function was found numerically,
and it was observed that the fit is better if one uses 
a  power-law form rather than  a Gaussian \cite{Frederico:2001qy,Pauli:2001uf}.
Furthermore, the power-law wave functions were  used some time ago 
in models for the nucleon form factors \cite{schlumpf}.  
In the present paper, we show that a simple power-law {\it Ansatz}  for 
the pion LC wave function  allows one to obtain 
explicit analytic expressions for the form factor
and generalized parton distributions.
To make our presentation self-contained,
in Section II we remind  basic 
information about   generalized parton 
distributions which is used in the following Sections.
In Section III, as a starting example, 
we consider the model with the Gaussian dependence
of the LC wave functions  
on the transverse momentum. 
In Section IV, 
we specify the explicit ``toy'' model expression 
for the effective pion wave function, which is then used 
in Section V  to   derive a parametric representation 
for the pion form factor. 
We show that  the two parameters of this simple  model,
the constituent quark mass and the wave function width, 
can be easily adjusted to provide a curve close to existing experimental 
data.
 In Section VI, we analyze  the asymptotic
 large-$Q^2$ behavior 
 of the pion form factor. We consider both 
 the massive $m \neq 0$ and massless $m=0$ cases.
 We show that, in the latter case, the pion form factor 
 in our model with a power-law wave function
 $\psi (x,k_{\perp}) \sim 1/[\sqrt{x(1-x)} (1 +
 k_{\perp}^2/(\lambda^2 x(1-x))^n]$ has the same asymptotic behavior
 $F_\pi (Q^2) \sim 1/Q^2$  for any power $n$.
 Though this  behavior is generated by the soft (Feynman) 
 mechanism, it formally coincides 
 with the quark counting law dictated 
 by the hard one-gluon exchange mechanism.
 We show that the $1/Q^2$ behavior of the soft contribution
 is related to the fact that the parton distribution 
 $f(x)$ in the massless case does not vanish  for $x=1$.
 In massive case, $f(x)\to 0 $ as $x \to 1$ and 
 the ultimate  asymptotic behavior is 
  $\ln (Q^2 / \lambda^2) /Q^4$.
  However, for a wide range of accessible $Q^2$, the curve 
  mimics the $1/Q^2$ behavior. 
   In Section VII,  we note that 
our  parametric representation for the form factor 
has the form of the reduction relation 
connecting the pion form factor and the double distribution  (DD) 
$F(x,y;t)$. 
   The DD obtained in this way has correct spectral and symmetry
properties. Moreover, it has the factorized structure 
proposed in Ref. \cite{Radyushkin:1998es}:  it looks like  
a distribution amplitude with respect to 
the $y$ variable and like a parton density
with respect to the $x$ variable. 
It also provides a nontrivial example of the interplay between
$x$, $y$ and $t$ dependence 
 of DDs. With an explicit model for DDs at hand, one can calculate 
 the relevant skewed distributions: 
 the nonforward parton distribution ${\cal F}_\zeta (X;t)$
 or Ji's off-forward parton distribution 
 $H(x,\xi,t)$, see Section VIII. 
In the simple toy model that we use  the pion is treated as 
  an effectively two-body system,
  which is not very realistic: 
  one may expect that the parton densities at
  small $x$ are  affected by many-body
  components. Indeed, the valence parton density
  obtained in our model differs rather strongly
  from the phenomenologically established form.
  In Section IX, we propose to fix this deficiency 
  by adopting a model with a more realistic 
  $x$-profile at $t=0$, but preserving the 
  analytic structure of the interplay 
  between $x,y$ and $t$ dependence generated 
  by the power-law {\it Ansatz}.
  We show that by slightly changing 
  the quark mass and the wave function width parameter
  it is still possible to get a good description of 
  the pion form factor data.
  We present  SPDs  ${\cal F}_\zeta (X; t)$ obtained from the 
  ``realistic'' DD.
  In particular, we show that
  in the ``soft pion limit'', $\zeta =1, t=0$,
  the isovector part of the ``realistic'' SPD
  has the shape close to the asymptotic form
  of the pion distribution amplitude.
  In  Appendix A, to demonstrate 
  that the variables $x,y$ of the parametric
  representation for the form factor 
  indeed have the meaning of 
  the variables of double distributions, we give a 
  covariant derivation of the toy model DD in a scalar model. 
  In Appendix B, we discuss the structure of model SPDs
  in the impact parameter  representation.
  In particular, we show how one can 
  use  superpositions of power-law  DDs to build 
  models for SPDs satisfying positivity bounds. 
  In Appendix C, using again the toy scalar model,
  we briefly show how 
  one can use our approach to build the models 
  for two-pion distribution amplitudes that appear in 
  $\gamma^* \gamma \to \pi \pi$  reaction, which can be treated 
  as the crossed-channel process to deeply virtual Compton scattering. 
  Our conclusions are formulated 
  in Section X. 
  
  Summarizing, 
  in this paper we  construct 
power-law     models of the 
$C$-odd   double distributions $F (x,y;t)$ for  the pion
   and the relevant  
  skewed parton distributions  ${\cal F}_\zeta (X;t)$. 
  By construction,  the model GPDs satisfy such  important constraints
  as reduction relations to usual parton densities and form factors,
  they have correct spectral and 
  polynomiality properties, thus providing
  a  model that can be used in phenomenological
  applications. For the simplified scalar case,
  we also build the models that  automatically satisfy 
  the positivity constraints.

\vskip .2in

\section{Basics of Generalized Parton Distributions}

Generalized parton distributions   parametrize nonforward 
matrix elements of composite operators. 
To define the leading-twist 
GPDs for the pion, we start with 
\begin{eqnarray}
 i^n \langle P-r/2 \, | \, \bar \psi_a \{ \gamma_{\mu} \stackrel{\leftrightarrow}{D}_{\mu_1}  \ldots
\stackrel{\leftrightarrow}{D}_{\mu_n} \} \,  \psi_a \, | \, P+r/2 \rangle
= 2  \sum_{k=0}^n \frac{n!}{2^k k! (n-k)!} \, A_{nk}^{(a)}(t) \, \{ P_{\mu} P_{\mu_1}  \ldots
P_{\mu_{n-k} } 
r_{\mu_{n-k+1}} \ldots  r_{\mu_n} \} \nonumber \\ 
+ \frac1{2^n} D_n^{(a)}(t) \{ r_{\mu} r_{\mu_1}  \ldots
 r_{\mu_n} \} \ , 
\label{DDmom} 
\end{eqnarray}
where $\stackrel{\leftrightarrow}{D} = (\stackrel{\rightarrow}{D} -\stackrel{\leftarrow}{D})/2$,  
 $\{ \ldots \}$ denotes the symmetric-traceless part of a tensor,
  $a$  numerates quark flavors and the quark fields are taken at the origin.
 Compared to the more familiar case of forward matrix 
elements defining the usual parton densities,
we have two 4-vectors, $P$ and $r$, both of which  can be used 
to build the tensor structure of the right hand side of 
Eq. (\ref{DDmom}). 
The index $k$ specifies how many times 
the vector $r$ appears in a particular term
of the sum.
Incorporating hermiticity 
properties of the local operators and time-reversal invariance,
one can show  \cite{Ji:1998pc}
that $k$ 
 is even. 
Now one can define 
double distributions  $f(\beta,\alpha;t)$ 
as functions generating $A_{nk}(t)$ through   its 
$\beta^{n-k} \alpha^k$ moments 
\begin{equation}
{\{1\pm(-1)^n\}} A_{nk}^{(a)}(t) = 
\int_{-1}^1  d\beta \int_{-1+|\beta|}^{1-|\beta|} 
\beta^{n-k} \alpha^k f^{\mp}_a(\beta,\alpha;t) \, d\alpha \ .
\label{DDnk}
\end{equation}
The spectral property $|\beta|+|\alpha|  \leq 1$ can be 
proved for  any relevant 
diagram of perturbation theory \cite{Muller:1998fv,Radyushkin:1997ki}.

As usual, the Mellin moments define two functions:
$f_a^{-}(\beta,\alpha;t)$ corresponds to even $n$ while 
$f_a^{+}(\beta,\alpha;t)$ to odd $n$. 
They both  are even functions of $\alpha$.
With respect to $\beta$,  $f_a^{-}(\beta,\alpha;t)$ is even 
while $f_a^{+}(\beta,\alpha;t)$ is odd.
For $\beta>0$, one can write $f_a^{-}(\beta,\alpha;t)$ 
as the difference $f_a(\beta,\alpha;t)-f_{\bar a}(\beta,\alpha;t)$
of quark and antiquark distributions ({\it i.e.},
$f_a^{-}(\beta,\alpha;t)$ corresponds to a valence 
quark distribution: $f_a^{-} = f_a^{\rm val}$ ) and $f_a^{+}(\beta,\alpha;t)$ 
as their sum $f_a(\beta,\alpha;t)+f_{\bar a}(\beta,\alpha;t)$.
The Polyakov-Weiss $D$-term \cite{Polyakov:1999gs}
is defined as the 
function $D_a(\alpha;t)$ whose $\alpha^{n}$ moments 
give the $D_n^{(a)}(t)$ coefficients.   
The latter are nonzero only for odd $n$, hence $D_a(\alpha;t)$
is an odd function of $\alpha$.

We stress that this definition 
of  double distributions 
is absolutely Lorentz invariant:
it does not require  reference to any particular
frame. Moreover, the mutual orientation and relative size 
of two momenta $P$ and $r$ are  arbitrary.
If, in some particular frame, the space part of the  momentum $P$  is 
oriented in the (longitudinal) $x_3$-direction, the 4-momentum $r$ 
may also have a nonzero longitudinal component,
but it may  be purely transverse as well,
having nonzero components  in 
the transverse $x_1,x_2$--plane only. 
The double distributions $f(\beta,\alpha;t)$ 
parametrizing the nonforward matrix element are Lorentz invariant objects
and they are  the same
in all cases.

Usually, to  extract the symmetric-traceless part
of a tensor $O_{\mu \mu_1 \ldots  \mu_n}$, it is  multiplied 
by $z^{\mu} z^{\mu_1}  \ldots z^{\mu_n}$, where
$z^{\mu}$ is a lightlike vector $z^2=0$. 
This trick corresponds to a  projection 
of Eq. (\ref{DDmom}): 
\begin{equation}
 \langle P-r/2 \, | \, \bar \psi_a 
 \hat z (i z\stackrel{\leftrightarrow}{D})^n \psi_a \, | \, P+r/2 \rangle
= 2 (Pz) \sum_{k=0}^n \frac{n!}{2^kk! (n-k)!} \, A_{nk}^{(a)}(t) \, (Pz)^{n-k} (rz)^k 
+ \frac1{2^n} D_n^{(a)}(t) (rz)^{n+1}\ , 
\label{DDz}
\end{equation}
(where  $\hat z \equiv \gamma_{\mu}z^{\mu}$).
The direction of $z$ is arbitrary, but, to  access 
all the coefficients $A_{nk}(t)$, one should have 
 both $(Pz)\neq 0$ and $(rz)\neq 0$. 
In particular, if $z$ has only the minus light-cone
component, both $P^+$ and $r^+$ should be nonzero
to make all the coefficients $A_{nk}$ visible.
Such a situation is characteristic for deeply virtual 
Compton scattering (DVCS) where the momentum transfer
$r$ must have a nonzero longitudinal component. 
  To study DVCS, it is  convenient  to treat the ratio
  $\xi= r^+/2P^+$ as an independent 
  variable and define off-forward  parton distributions 
  $H(\tilde x,\xi;t)$ \cite{Ji:1996ek}. 
 To this end, one introduces the functions 
\begin{equation}
{\cal M}_n^{(a)} (\xi;t) =  \sum_{k=0}^n \frac{n!} {k! (n-k)!}\, A_{nk}^{(a)}(t)\, \xi^k + D_n^{(a)}(t) \xi^{n+1} 
\end{equation}
and declares ${\cal M}_n^{(a)} (\xi;t)$ to be the moments of  $H_a(\tilde x,\xi;t)$:
\begin{equation}
{\{1\pm(-1)^n\}} {\cal M}_n^{(a)} (\xi;t)= \int_{-1}^1 \tilde x^n \, H^{\mp}_a(\tilde x,\xi;t) \, d\tilde x \ .
\end{equation}
These definitions provide a formal relation 
between $H(\tilde x,\xi;t)$
 and $ f(\beta,\alpha;t)$ 
\bea{OFPDa}
 H^{\pm}_a(\tilde x,\xi;t) = \int_{-1}^1 d\beta \int_{-1+|\beta|}^{1-|\beta|} f^{\pm}_a(\beta,\alpha;t)  \, 
  \delta (\tilde x-\beta -\xi \alpha) \, d\alpha \  + (1\pm 1) {\rm sign} (\xi) \, D_a(\tilde x/\xi;t).
   \end{eqnarray}

Combining Eqs.(\ref{DDnk}) and (\ref{DDz}) gives the definition 
of DDs through the parametrization of  nonforward 
matrix elements of nonlocal light cone operators
\begin{eqnarray}
 \langle P-r/2 \, | \, \bar \psi_a (-z/2) \hat z  \psi_a (z/2)  \, | \, P+r/2 \rangle|_{z^2=0} 
&=&  (Pz)  \, \int_{-1}^1  d\beta \int_{-1+|\beta|}^{1-|\beta|} 
e^{-i\beta(Pz)-i\alpha(rz)/2} \, \Bigl (f_a^+(\beta,\alpha;t) + f_a^-(\beta,\alpha;t)\Bigr )\, d\alpha 
\nonumber \\ &+& (rz) \int_{-1}^1 e^{-i\alpha(rz)/2} D_a(\alpha;t) \, d\alpha \ .
\label{DDnonloc}
\end{eqnarray}

Using the symmetry  of $f_a^{\pm}(\beta,\alpha;t)$ with respect to $\beta$ and $\alpha$, 
one can rewrite this representation
in terms of quark and antiquark DDs taken for positive $\beta$  only
\begin{eqnarray}
 \langle P-r/2 \, | \, \bar \psi_a (-z/2) \hat z  \psi_a (z/2)  \, | \, P&+&r/2 \rangle|_{z^2=0} 
= 2(Pz)  \, \int_{0}^1  d\beta \int_{-1+\beta}^{1-\beta} 
 \nonumber \\ 
&\times& \Bigl (f_a(\beta,\alpha;t)e^{-i\beta(Pz)-i\alpha(rz)/2}
 - f_{\bar a} (\beta,\alpha;t)e^{i\beta(Pz)+i\alpha(rz)/2}\Bigr )\, d\alpha 
\nonumber \\ 
&+& (rz) \int_{-1}^1 e^{-i\alpha(rz)/2} D_a(\alpha;t) \, d\alpha \ .
\label{DDnonloc2}
\end{eqnarray}

In the forward limit,  $r=0$, the left hand side of Eq. (\ref{DDnonloc2}) 
coincides with the matrix element defining the usual parton densities
$f_{a,\bar a}(x)$. This gives the reduction relations 
\begin{equation}
 \int_{-1+x}^{1-x}  f_{a,\bar a} (x,\alpha;t=0) \, d\alpha = f_{a,\bar a} (x) \   \   \  {\rm and} \   \   \ 
 H_{a,\bar a}(x, \xi=0;t=0) = f_{a,\bar a} (x) \  .
\end{equation} 
On the other hand, keeping $r\neq 0$ but 
taking $n=0$ in  Eq. (\ref{DDmom}) 
one deals with the matrix element of the vector current
which defines the $a$-component $F_a(t)$ of the relevant  form factor.
The reduction relations connecting GPDs  with   form factors 
result from 
\begin{equation}
F_a(t) =A_{00}^{(a)}(t) = {\cal M}_0^{(a)}(t)
 \end{equation}
 and are given by expressions 
\begin{equation}
\int_0^1 d\beta \int_0^{1-\beta} \Bigl (f_a(\beta,\alpha;t) - f_{\bar a}(\beta,\alpha;t) \Bigr ) 
 \, d\alpha = F_a(t) \  \  \ \ , \  \    \  \
\int_0^1  \Bigl (H_a(\tilde x,\xi;t) - H_{\bar a}(\tilde x,\xi;t) \Bigr) \, d\tilde x = F_a(t)  
 \end{equation}
containing only  the valence quark combinations  
$f_a^{\rm val} = f_a - f_{\bar a}$ and 
$ H^{\rm val}_a =  H_a - H_{\bar a}$.  

The representation (\ref{DDnonloc2})  has the structure of a  
plane wave decomposition, which provides the parton interpretation
of DDs: the  quarks  
carry the momentum $\beta P+(1+\alpha)r/2$  originating both
from the average  momentum $P$ and the momentum transfer $r$.
Another possibility (which is more convenient 
in applications involving light-cone wave functions)
is to write the momenta of quarks  as $xp_1+yr$,
{\it i.e.,} in  terms of $r$ and the 
original hadron momentum $p_1=P+r/2$. The new variables $x,y$ 
are expressed through $\beta,\alpha$
by $x=\beta, y= (1+\alpha - \beta)/2$.  
The resulting DDs $F_{a,\bar a} (x,y;t)$ ``live'' on the triangle
$0\leq x,y,x+y \leq 1$. Since $f(\beta,\alpha;t)$ are even functions of $\alpha$,
the DDs $F (x,y;t)$  are symmetric with respect to 
$y\to 1-x-y$ transformation ({\it ``Munich'' symmetry} \cite{Mankiewicz:1997uy}). 
 For light-cone dominated processes, like DVCS, only the plus component 
 $xp_1^+ + yr^+$ is essential. Defining the skewedness 
 parameter  $ \zeta = r^+/p_1^+$, we  introduce
 nonforward parton distributions 
 \cite{Radyushkin:1997ki} 
\bea{NFPDa}
  {\cal F}_{\zeta}^{a,\bar a}  (X,t) = \int_0^1 dx \int_0^{1-x}  F^{a,\bar a} (x,y;t) \, 
  \delta (X -x -\zeta y) \, dy \ .
   \end{eqnarray}
These  distributions  are related to  usual parton densities by 
\bea{redf}
\int_0^{1-x}  F_{a,\bar a}(x,y;t=0) \, dy = f_{a,\bar a}(x) \   \   \  ,  \   \   \ 
 {\cal F}_{\zeta=0}^{a, \bar a} (X;t=0)= f_{a, \bar a}(X)  
\end{eqnarray} 
and to form factors by  
\bea{redF}
\int_0^1 dx \int_0^{1-x} F_a^{\rm val}(x,y;t) \, dy = F_a(t) \  \  \ \ , \  \    \  \
\int_0^1  {\cal F}_{\zeta}^{a, \rm val} (X,t) \, dX = F_a(t)  \ .
 \end{eqnarray}

Note that the double distributions  $F(x,y;t)$  
are integrated over $y$ in both of the above reduction 
relations. Thus, it makes sense to introduce  intermediate  
functions
\begin{equation}
{\cal F}_{a, \bar a} (x,t) = \int_0^{1-x} F_{a, \bar a}(x,y;t) \, dy   \equiv 
{\cal F}_{\zeta=0} (x;t) \ . 
 \end{equation}
They  satisfy simpler reduction relations
\begin{equation}
{\cal F}_{a, \bar a} (x,t=0) = f_{a, \bar a} (x) \ \ \ {\rm and} \   \   \ 
\int_0^1  {\cal F}_a^{\rm val}(x,t) \, dx = F_a(t)  \  .
\end{equation}
Thus, the functions  ${\cal F}(x,t)$ are hybrids of  form factors $F(t)$ 
and  usual parton densities $f(x)$,
that is why  we  call  them {\it nonforward parton densities} (NDs)
\cite{Radyushkin:1998rt}.

\section{Gaussian Wave Function and Nonforward Parton Densities}

The concept  of NDs  can be easily 
illustrated within 
 the framework of the light-cone formalism. 
Consider  
a two-body bound state  whose 
lowest Fock component is described by a light-cone   
wave function $\psi(x,k_{\perp})$.
In  a frame where the momentum transfer $r$  is purely 
transverse  $r=  r_{\perp}$, one can write  
the two-body contribution into the form factor  
as \cite{Drell:1969km}
\begin{equation}
F^{(2b)} (t) = \int_0^1 \,  dx \, \int \, 
\psi^* ( x, k_{\perp}+ (1-x)  r_{\perp}) \,  
\psi (x, k_{\perp}) 
\, {{d^2 k_{\perp}}}  \equiv \int_0^1 \, {\cal F}^{(2b)} (x,t)  dx \, , 
\end{equation}  
\noindent where  $F^{(2b)} (x,t)$ is 
the  2-body contribution into the nonforward  parton density
 \begin{equation} 
 {\cal F}^{(2b)} (x,t) = \int  \, 
 \psi^* ( x, k_{\perp}+ (1-x) r_{\perp}) \,
\psi (x, k_{\perp}) \, 
{{d^2 k_{\perp}}} \ .
\end{equation}  
Adding contributions from higher Fock components,  one obtains 
the total ND ${\cal F} (x,t)$ whose integral over $x$ gives the 
form factor $F(t)$
of the
bound state. As discussed in the previous section, 
at zero momentum transfer  ${\cal F} (x,t)$ reduces to 
the usual valence parton density   
$f(x)= {\cal F} (x;t=0) $. Furthermore, there is the usual form factor
normalization 
condition  
$F(t=0)=1$. 
Finally, for the  valence quark distributions, the integral of $f(x)$ 
over $x$ is 1.  
 These
conditions   are satisfied  in the simplest way
by the factorized {\it Ansatz} ${\cal F} (x,t) = f(x)F(t) $,
in which there is no interplay between $x$ and $t$ dependence
of ${\cal F} (x,t) $. 
One may expect that in reality the situation is more complicated.
Consider  a wave function with a Gaussian dependence on the transverse momentum $k_{\perp}$
(cf. \cite{huang})    
\begin{equation}\psi (x,k_{\perp}) =  \varphi(x) 
 e^{-k^2_{\perp}/2x (1-x) \Lambda^2}  \label{11} 
 \end{equation}
 (note, that $ k^2_{\perp}/4x (1-x)$ is essentially $\vec k^2$ written in 
the light-cone variables $x,k_{\perp}$). 
Taking the Gaussian integral  over $x,k_{\perp}$ we get 
\begin{equation}
{\cal F}^{(2b)} (x,t) = f^{(2b)}(x) e^{(1-x ) t /4 x \Lambda^2 } \, , \label{8}
\end{equation}
where 
\begin{equation}
f^{(2b)}(x) =   \pi
x (1-x) \, \Lambda^2  \, \varphi^2(x) 
= {\cal F}^{(tb)} (x,t=0)
\end{equation}
is the two-body part of the relevant parton density $f(x)$.
To get the total
result for either  usual $f(x)$
or 
nonforward parton densities   ${\cal F}(x,t)$,
one should 
add the contributions due to  higher Fock components.
These contributions
are not small, {\it e.g.,} with the Gaussian {\it Ansatz} the  
valence $\bar d u$  contribution
into the normalization of the $\pi^+$  form factor 
for  $t=0$ is about 25\% \cite{huang}. 
The problem is that we do not have a formalism providing  explicit expressions
for an infinite tower of light-cone wave functions. 
However, the parton densities $f(x)$ are known from experiment.
In this situation, one can   treat  Eq.(\ref{8}) as a guide
for  fixing interplay between  $x$ and $t$ dependence
of NDs and 
 model them by 
\begin{equation}
{\cal F}^a(x,t) = f_a(x) e^{\bar x t /4 x \Lambda^2 }  \,  . \label{13}
\end{equation}
The functions  $f_a(x)$  here are the 
usual valence $a$-quark parton densities. One can take them  
from existing  parametrizations of parton densities like GRV, MRS, CTEQ, {\it etc.} 
This model (originally proposed in Ref. \cite{Barone:ej}) 
was succesfully applied in Ref. \cite{Radyushkin:1998rt} 
to describe the proton form factor $F_1(t)$
in a wide region $1 < -t < 10$\,GeV$^2$  
of momentum transfer by fitting the only parameter
$\Lambda^2$ characterizing the effective proton size.

\section{Power-Law Wave Functions}

 GPDs give the most general 
parametrization of nonforward matrix elements.
Furthermore, both of them, DDs $F(x,y;t)$ and SPDs ${\cal F}_{\zeta} (X;t)$
are functions of 3 variables: in addition to the invariant 
momentum transfer $t$  they depend on two ``longitudinal'' variables
$x,y$ or $X, \zeta$. However, the Gaussian model
of the previous section gives a representation
for  the form factor in terms of  a one-dimensional  $x$-integral
of the function ${\cal F} (x,t)$  depending on  only 2 variables, $x$ and $t$.
One may suspect that  the Gaussian {\it Ansatz} is a  degenerate  case
failing to reveal the richer structure present in  
more  general situations.
In what follows, our goal is to study
a  model based on power-law wave functions. 
As we will see, though this model is more complicated, we are  still 
able  to get most of the results in  analytic form,
which allows us to use it for building 
nontrivial  {\it Ans{\"a}tze} for generalized parton distributions.

The  $q \bar q$ wave function of the pion found numerically in  
\cite{Frederico:2001qy} was  
parametrized analytically by a power law  fit  
\beq{eq:ff}
   \varphi(\vec k) \sim \left(\frac{1}{1+k^2/\Lambda^2}\right)^{\kappa}
, 
\end{equation}
with $\kappa \sim 2$ rather than 
by a Gaussian $\varphi(\vec k) \sim \exp(-k^2/\Lambda_G^2) $.
Here $k^2= k_z^2 +k_{\perp}^2$ is the square of the relative 3-momentum 
and $\Lambda$ is the parameter characterizing 
the width of the $k^2$-distribution, {\it i.e.,} the size of the system.
In the case of equal quark masses, there is a simple  
relation \cite{Terentev:jk} between the usual variables
$(k_z,k_\perp)$  and the infinite momentum frame (IMF) 
variables $(x, k_\perp)$ 
\bea{xz}
x={1\over 2} \left [1+{k_z\over {\sqrt {m^2+k_z^2+k_\perp^2}}} \right ],
\end{eqnarray}
where $m$ is the effective quark mass.
The relation between $ \varphi(\vec k)$ and the IMF wave function 
$\psi(x,k_\perp)$ is given  by
\bea{phipsi}
\psi(x,k_\perp)=\varphi({\vec k})\, \frac{(1+{ k}^2 /  m^2)^{1/4}}{
\sqrt{x(1-x)}}.
\end{eqnarray}
For light quarks, one may expect that the size parameter 
 $\Lambda$ is close to the effective quark mass $m$.
 Then   the factor $(1+{ k}^2 / m^2)^{1/4}$ 
can be essentially absorbed into a redefinition of the power $\kappa$,
whose precise value, in fact,  is not critical for our purposes.
Thus, in what follows,   we will consider a 
 simplified power-law IMF wave function
\bea{psi}
\psi(x,k_\perp)= \frac{N}{\sqrt{x(1-x)}[a+bk_\perp^2]^2},
\end{eqnarray}
where $a+bk_\perp^2$ is the IMF version of $(1+k^2/\Lambda^2)$ with 
\bea{a,b}
a=1+{s^2 (x-{1\over 2})^2\over {x(1-x)}},~~~~~~~~~~~~ b={s^2\over {4 m^2
x(1-x)}}\, , ~~~~~~s={m\over \Lambda}\, ,
\end{eqnarray}
and $N$ is the normalization
constant. For the 2-body Fock component,
it can be fixed from the requirement that 
the integral of $\psi(x,k_\perp)$ over $x$ and 
$k_\perp$  should give the pion decay constant
\begin{equation}
{ \sqrt \frac3{2\pi^3} } 
\int \,  \psi^{(2b)}(x,\ k_{\!\perp}) \, dx\, d^2  k_{\!\perp}
=f_{\pi} \ .
   \end{equation} 
As noted in the previous section, the knowledge of the 2-body wave function is not 
sufficient  to calculate the pion form factor. To get it, 
we should add  the contribution from all higher Fock  components.
Just like in the case of the Gaussian wave function,
the two-body component is responsible only for  some 
portion of ``1'' in the normalization condition
$F(0) = 1$
\cite{huang,Pauli:2001uf}. 
Again, the structure of  higher Fock components
can be only guessed. 
 To avoid making  too many  guesses,
 we will  analyze the simplest ``one-guess'' model in which 
a single  two-body-like  function  
$\psi(x,k_\perp)$ (\ref{psi}) imitates
the contribution of all  Fock components 
into the pion form factor. Thus, we take  
\beq{DY}
   F (Q^2) =
  \int \, 
   \psi(x, k_{\!\perp}+(1-x) q _{\!\perp})\, 
   \psi(x,\ k_{\!\perp}) \, dx\, d^2  k_{\!\perp}  \ , 
   \end{equation} 
and normalize the  effective wave function 
 $\psi(x,k_\perp)$   by 
\bea{norma}
\int  {\mid \psi(x,k_\perp) \mid }^2 \, dx \, d^2 k_\perp=1 \ . 
\end{eqnarray}
This   condition (\ref{norma}) gives an explicit expression 
for the normalization constant $N$ of the effective wave function:
\bea{N2}
N^2={3\over {4\pi}} \left ( {s\over m} \right )^2  {1\over {A(s)}} \ , 
\end{eqnarray}
where 
\bea{A(s)}
A(s)=\int_0^1  
 {dx \over a^3} = \int_0^1  {(1-z^2)^3\over [1-(1-s^2)z^2]^3} \, dz.
\end{eqnarray}

Before proceeding further, we would like to make it clear that 
substituting the total contribution 
of higher Fock components by a 2-body type term is just a toy model,
and we do not expect it to adequately describe 
all the aspects  of the pion structure. 
In particular, the total parton density in the toy model
has the same ($x \to 1-x$ symmetric) 
shape as its 2-body part, and it vanishes 
at $x=0$. 
One would expect, however,  that
the contributions of higher Fock components are
shifted to smaller and smaller $x$ values
producing eventually the experimentally observed 
$\sim 1/\sqrt{x}$ behavior. 
We do not know how much 
 each term of the infinite tower
of Fock components contributes to
the parton density, but we know (from experiment)  
what is  the total result. 
 Thus,  our ultimate strategy,  
just like in the case of the Gaussian wave function, 
is to calculate GPDs in the toy model, identify the factor
corresponding to the usual parton density 
and substitute it by the experimental one. 
On the other hand, one may expect that 
form factor, being an integral of the relevant GPD,
should not be too  sensitive to the details of its
$x$-dependence, at least in some range of momentum transfer $t$.  
 Thus, we study first the form factor in our  toy model.
 We show that, despite its  crudeness, the toy
 model  can easily fit the 
form factor data by adjusting the two parameters of the model.
Then we incorporate the main advantage of
the toy model, the possibility 
to do  calculations analytically, 
and obtain the  representation for the form factor in terms
of DDs. Finally, we ``correct'' the latter in such a way  that, after
integration, they 
produce  experimental parton densities. 
We also show that this model gives DDs with a 
nontrivial  ``profile''  dependence on the $y$ variable.


\section{Form Factor in Toy Model} 

The $k_\perp$ integral in the  expression  for the form factor
\bea{Fint}
F(Q^2) = N^2 \int { dx \, d^2 k_\perp \over{x(1-x)}
[a+b(k_\perp+(1-x)q_\perp)^2]^2  [a+bk_\perp^2]^2}
\end{eqnarray}  
can be done using either the Feynman parameters
or the Schwinger  $\alpha$-representation method briefly 
described below. 
To  this end, we  use
\bea{alphar}
{1\over A^{\kappa}}= {1\over \Gamma(\kappa)} \int_0^\infty 
 \alpha^{\kappa -1}  \, e^{-\alpha A} \, d
\alpha \ , 
\end{eqnarray}
where $\kappa=2$ in our case. 
After calculating the Gaussian integral
over $k_{\perp}$, we arrive at the representation 
for the form factor in  terms of two parameters $\alpha_1 $ and $\alpha_2$
\bea{a1a2}
F(Q^2) ={ \pi N^2 }  \int_0^\infty \! \int_0^\infty  \! \frac{\alpha_1
d \alpha_1 \,  \alpha_2
d \alpha_2}{(\alpha_1
+\alpha_2)} \int_0^1  
e^{-a(\alpha_1+\alpha_2)}e^{-b(1-x)^2 Q^2 {\alpha_1 \alpha_2 \over {\alpha_1
+\alpha_2}}}{dx \over {b x(1-x)}} \ ,
\end{eqnarray}
where  $Q^2= q_\perp^2$.
Changing  the variables 
\bea{scaling}
 \alpha_1+\alpha_2=\lambda,~~~~\alpha_1=\gamma \lambda,~~~~ 
 \alpha_2 = (1-\gamma) \lambda, ~~~~ d \alpha_1 d \alpha_2 = \lambda d \lambda
 d \gamma  \ , 
\end{eqnarray}
 we  obtain   the parametric representation 
\bea{Fbla}
F(Q^2) = { \pi N^2 } \int_0^1 {dx \over {b x(1-x)}}
\int_0^1 d\gamma \,  \gamma (1-\gamma ) \int_0^\infty \lambda^2\, d\lambda \, 
 e^{-[a\lambda +b\lambda (1-x)^2 \gamma (1-\gamma ) Q^2]}  \ .
\end{eqnarray}
Integration over $\lambda$   is easily performed to give 
\bea{Fxbe}
F (Q^2)  &=&2 \pi N^2 \int_0^1 {dx \over {bx(1-x)}}
\int_0^1 d\gamma \, { \gamma (1-\gamma )\over {[a+b(1-x)^2
\gamma (1-\gamma ) Q^2}]^3} \ .
\end{eqnarray}
 Incorporating the normalization condition, we get the final result 
\bea{Ffin}
F_{\pi}(Q^2) &=&  {1 \over A(s)} \int_0^1 dx 
\int_0^1 d\gamma   \, {6 \gamma (1- \gamma)  \over {[a+b(1-x)^2
\gamma (1- \gamma) Q^2}]^3} \ .
\end{eqnarray}
By construction, the form factor has the correct 
value  at $Q^2=0$. However, its slope at this point 
depends on the values of the model parameters 
$m$ and $s$. To obtain the analytic expression for the slope 
we note that, in the small $Q^2$ limit,  
one can  expand the denominator of the $\gamma $
integration 
\bea{FsmallQ}
F_\pi (Q^2) \mid_{Q^2 \rightarrow 0} &=&   {6 \over A(s)} \int_0^1 
{dx \over {a^3}} \int_0^1  d\gamma   \, \gamma (1- \gamma)  \, 
\Big [1- {3b\over a}   \gamma (1- \gamma) Q^2
(1-x)^2 \nonumber \\ &&~~~~~~~~~
 +{6 b^2\over a^2}(\gamma (1- \gamma))^2 Q^4 (1-x)^4 + \ldots \Big ].
\end{eqnarray}
Using  the normalization condition (\ref{norma}) for  the wave function, 
taking the  integrals
over $\gamma$ and introducing  
the variable $z= 2x-1$ (like in Eq. (\ref{A(s)})) we
obtain 
\bea{FsmallQ2}
F_\pi (Q^2) {\mid}_{Q^2 \rightarrow 0}=1-{3\over 20}{Q^2\over m^2}s^2 {B(s)\over
A(s)}+{9\over 280}{Q^4\over m^4}s^4 {C(s)\over A(s)} + \ldots \ ,
\end{eqnarray}
where 
\bea{B(s)}
B(s)=\int_0^1  {(1+z^2)(1-z^2)^3\over (1-(1-s^2)z^2)^4} \, dz
\end{eqnarray}
and 
\bea{C(s)}  
C(s)=
\int_0^1 {(1+z^4+6z^2)(1-z^2)^3\over (1-(1-s^2)z^2)^5}\, dz \ .
\end{eqnarray}

\begin{figure}[t]
\hspace{4cm}
\mbox{
   \epsfxsize=6cm
 \epsfysize=5cm
 \hspace{0cm}
  \epsffile{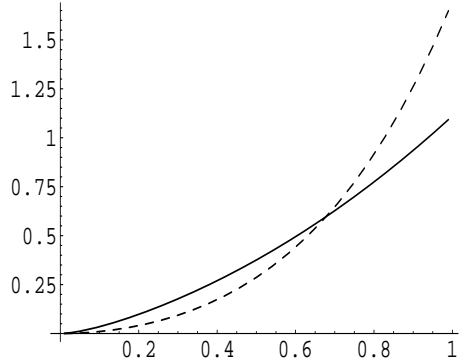}  } \hspace{0cm}
\caption{\label{A-B-C}
Combinations $s^2 B(s)/A(s)$ (solid) and $s^4 C(s)/A(s)$ (dashed)
as functions of the parameter $s$.
}
\end{figure}

The integrals $A(s)$, $B(s)$ and $C(s)$ can be calculated in
elementary functions,
though the results are rather lengthy. Fig. \ref{A-B-C}
 shows the plot of the  combinations 
$s^2 B(s)/A(s)$ and $s^4 C(s)/A(s)$ as  functions of the parameter $s$. 
The combination $s^2 B(s)/A(s)$ is monotonically increasing 
from zero to infinity. Hence, after choosing  the effective 
mass $m$ we can always find such a parameter $s$ that 
the slope $d F_\pi (Q^2) /dQ^2$  of the pion form factor at $Q^2=0$
has the experimental value $d F_\pi^{exp} (Q^2) /dQ^2 \approx 1/m_\rho ^2$
\cite{Amendolia:1984nz}.
For masses $m= 0.2\, , \, 0.3 $ and $0.4$  GeV,
the parameters $s$ fitting the pion charge radius are 0.56, 0.95   and 1.33,
respectively.

\begin{figure}[t]
\mbox{
   \epsfxsize=6cm
 \epsfysize=5cm
 \hspace{1cm}
  \epsffile {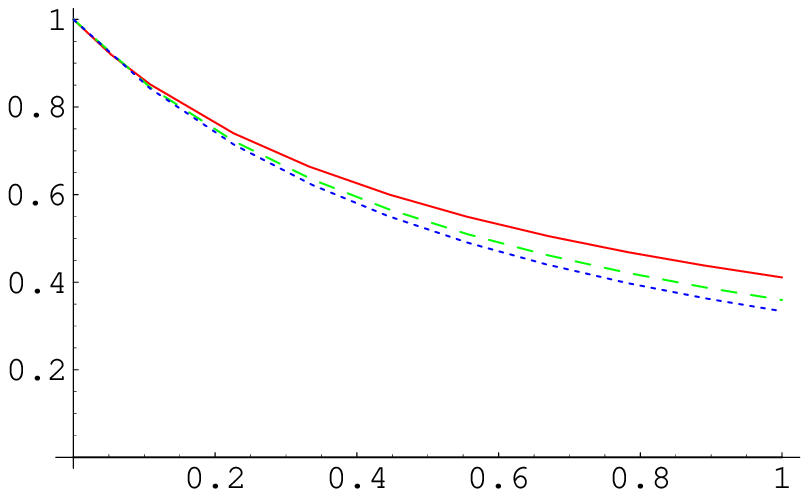} } \hspace{0cm}
\mbox{
   \epsfxsize=6cm
 \epsfysize=5cm
 \hspace{1cm}
  \epsffile {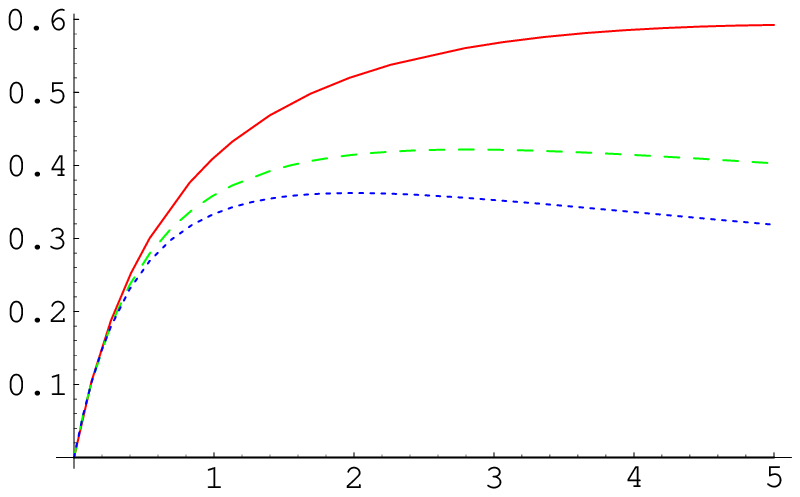} } \hspace{0cm}
  \vspace{3mm}
\caption{\label{FormFactor-orig}
Left: Form factor $F_\pi(Q^2)$ for three different parameterizations
of the wave function: with $m= 0.2$\,GeV (solid), $0.3$\,GeV (dashed)
and $0.4$\,GeV (dotted). Right: $Q^2 F_\pi(Q^2)$, with the same
wave functions as in the left.
}
\end{figure}

In Fig. \ref{FormFactor-orig}, left,
 we have plotted the form factor as a function of $Q^2$ in the
low $Q^2$ region $Q^2 <1$\,GeV$^2$ 
for these three different parametrizations of the wave function. 
Since they have the same slope at $Q^2=0$, 
the curves are rather  close to each other.
However, the difference between the curves
becomes more pronounced as $Q^2$ increases.
In Fig. 2, right,  the form factor calculated with these three 
effective wave functions is shown  in the $Q^2$ region up to
5 GeV$^2$ relevant to future experiments at Jefferson Lab. 
Among these three choices, 
the most close to existing experimental data \cite{Volmer:2000ek}
is the curve corresponding to $m=0.3$\,GeV and $s=0.95$.
\section{Asymptotic Behavior of the Pion Form factor}

According to  Fig. 2, right, in the accessible 
energy range $Q^2 \lesssim 5\,$GeV$^2$, 
the model curves show the behavior 
close to the $1/Q^2$   scaling. Since the mass scales 
involved are rather small: $m^2, \Lambda^2 \sim 
0.1\,$GeV$^2$,  one may think that the form factor is already 
in the asymptotic region. 

To  analyze    the  asymptotic behavior 
of   the form factor one can follow the
approach described in Ref. \cite{Brodsky:1989pv}. 
The basic idea is that  in the Drell-Yan formula
(\ref{DY}) we deal with an  overlap 
of two functions $\psi(x,k_\perp)$  
and $\psi(x,k_\perp - (1-x) q_\perp)$ whose 
$k_\perp$ arguments are separated 
by a gap   $(1-x)Q$ in   magnitude.
Furthermore, $\psi(x,k_\perp)$ rapidly decreases with $k_\perp$.
Hence, when $Q^2$ is large, the integral over $k_\perp$
in the form factor expression is dominated by two regions of phase space 
\cite{Brodsky:1989pv}:

1) $\mid k_\perp \mid \ll (1-x) Q$, where  $\psi(x, k_\perp)$ is large;

2) $\mid k_\perp +(1-x) q_\perp \mid \ll (1-x) Q$,  where
 $\psi(x, k_\perp+(1-x)q_\perp)$ is large.

 In the first case, $k_\perp  $ can
be neglected compared to $(1-x)q_\perp$ in the wave function.
The contribution from this region  is then approximated by
\bea{Fappr}
F_\pi^{(1)}(Q^2) \sim \int_0^1  \psi(x, (1-x)Q) \, dx \int 
\theta \bigl (\mid k_\perp \mid <  (1-x) Q \bigr ) 
\psi(x,k_\perp) \, d^2k_\perp.
\label{fpias} 
\end{eqnarray}
Since the wave function falls off rapidly at large transverse momenta,
the major contribution to the integral comes from the region
where  $\mid k_\perp \mid$
is much smaller than $(1-x)Q$,
and one may hope that the $k_\perp$ integral of 
$\psi(x,k_\perp)$ can be approximated by the
pion distribution amplitude $\varphi (x)$.  
The next statement usually made is that 
the large-$Q$ behavior of the function $\psi(x, (1-x)Q)$ 
is determined by the large-$k_\perp$  behavior  of $\psi(x,k_\perp)$
and, hence, the large-$Q$ behavior of the form factor
just repeats the large-$k_\perp$ behavior of the wave function;
{\it i.e.,} if $\psi(x,k_\perp) \sim 1/k_\perp^n$, then
$F \sim 1/Q^n$. Note, that the last statement is only true
if, after these substitutions, the integral over $x$ converges. 
However, it is easy to derive that, after the 
$k_\perp$ integration in Eq.(\ref{fpias}),
the remaining integrand  for the $x$-integration is proportional
to 
\beq{xint}
\frac{x^3 (1-x)^5 Q^2}{\left [ x (1-x) +s^2 \left (x- \frac12 \right )^2 +
(1-x)^2\frac{ Q^2}{4 \Lambda^2} \right ]^3} \ .
\end{equation}
 Neglecting the $x(1-x)$ and $s^2  (x- 1/2)^2$ terms compared 
 to $ (1-x)^2 Q^2/4 \Lambda^2$, one would get the integral $dx/(1-x)$ 
 logarithmically diverging in the $x \to 1 $ region.
 Of course, this approximation is only true 
 when  $ (1-x)^2 Q^2/4 \Lambda^2 \gg 1$ or $x \ll 1-2\Lambda/Q$.
 This cut-off converts the logarithmic divergence into $\ln (Q^2/\Lambda^2)$.
Hence,  the asymptotic behavior is $\sim \ln (Q^2/\Lambda^2) /Q^4$. 
This result can also be obtained from our representation 
(\ref{Ffin}) for the form factor, which we write now as 
 \bea{Fasxbe}
F_{\pi}(Q^2) &=& \frac{1}{A(s)}   \int_0^1 dx 
\int_0^1 d\gamma  \, {6\, \gamma (1- \gamma) x^3 (1-x)^3  
\over {\left [x (1-x) + {s^2(x-1/2)^2} + 
\gamma (1- \gamma) \frac{ (1-x)^2 Q^2}{4 \Lambda^2 } \right ]^3}} \ .
\end{eqnarray}
Again, there are two regions: $\gamma \ll 4 \Lambda^2 /[(1-x)^2 Q^2]$ 
and $1-\gamma  \ll 4 \Lambda^2 /[(1-x)^2 Q^2]$ producing the leading
large-$Q^2$ contribution  $\sim   \Lambda^4 /[(1-x)^2 Q^2]^2$. 
Combined with other $x$-dependent factors,
this gives the $dx/(1-x)$ divergence or, after a more
accurate calculation, the $\ln (Q^2/\Lambda^2) /Q^4$ asymptotic 
behavior.  This behavior is not yet  visible in the curves shown
in Fig. 2, right. The curves suggest, in fact, the $1/Q^2$ 
behavior. The slow approach to asymptopia 
can be traced to rather small numerical factor 
$\gamma (1- \gamma)/4 \sim 1/16$ accompanying
the $Q^2$ term. As a result, the effective 
scale governing the  $Q^2$-behavior is something like  
$16 \Lambda^2\sim 1.5\,$GeV$^2$ rather than simply $\Lambda^2$. 
Thus, the quark mass squared $m^2 \sim 0.1$\,GeV$^2$ is 
small compared to the effective scale, and it is worth
investigating what happens
when   quarks are  massless, 
 {\it  i.e.,} when  $s=0$. Then 
\bea{Fmasslessq}
\left. F_{\pi}(Q^2) \right |_{s=0} &=&   \int_0^1 dx 
\int_0^1 d\gamma  \, {6\, \gamma (1- \gamma) x^3   
\over {\left [x    + 
\gamma (1- \gamma) \frac{ (1-x) Q^2}{4 \Lambda^2 } \right ]^3}},
\end{eqnarray}
and situation drastically changes: the large-$Q^2$ behavior is dominated 
by integration over $1-x \ll 4\Lambda^2 /[\gamma (1- \gamma)Q^2]$ region.
The remaining integral over $\gamma$ has no singularities,
and we get  
  $F_{\pi}(Q^2)  \sim  {  \Lambda^2 /  Q^2}  $ 
 for the asymptotic behavior.
 Clearly, the asymptotic behavior 
 for massless quarks is governed by the soft 
 (or Feynman) mechanism.  One can easily check that the same 
 asymptotic power law
 $F_{\pi}(Q^2) \sim 1/Q^2$ holds for the 
 $\psi ( x, k_{\perp}) \sim 1/[\sqrt{x(1-x)} (1+bk_{\perp}^2)^{\kappa}]$ 
 wave functions with any power $\kappa$,
 and also for the exponential wave function
 $\psi ( x, k_{\perp}) \sim e^{-bk_{\perp}^2}/\sqrt{x(1-x)}$.
 This puzzling observation has a rather simple explanation:
 the valence parton  densities 
 in these models with massless quarks are constant: $f(x) =1$,
 and it is this singular $f(x)|_{x\to 1}  \to {\rm const} $
 behavior which is responsible for the $1/Q^2$ 
 contribution to the form factor.
 If we would ``correct'' the model so that
  $f(x)$ has a more realistic $\sim (1-x)$ behavior for $x$
  close to 1, the Feynman mechanism contribution would  
 have  a $1/Q^4$ asymptotic behavior.

 An efficient way  to obtain the asymptotic expansion 
 in powers and logarithms of $\Lambda^2 / Q^2$
is based on the Mellin representation for the 
denominator factor:
  \bea{Mellin}
   \left [x + \gamma (1-\gamma) (1-x) \frac{Q^2}{4 \Lambda^2} \right ]^{-3}
  &=&  \frac1{2 \pi i} \int_{-\delta - i \infty}^{ -\delta + i \infty} 
  \Gamma (-J) \Gamma (J + 3) \nonumber \\
  & & \ \ \ \ \times \gamma^J (1- \gamma)^J
  (1-x)^J x^{-J-3} \left ( {Q^2 \over 4\Lambda^2} \right )^J 
  \, dJ \, .
  \end{eqnarray} 
 Now, the $\gamma$ and $x$ integrals can be calculated
 in $\Gamma$-functions to give  
 \bea{FGamma}
F_{\pi}(Q^2) &=&  \frac1{2 \pi i} \int_{-\delta - i \infty}^{ -\delta + i \infty} 
6 \, { \Gamma (-J) \Gamma (1-J)\over (J+1) \Gamma (2J+5)}  \Gamma^2 (J + 2)
\Gamma^2 (J + 3) \left ( {Q^2 \over 4\Lambda^2} \right )^J 
  \, dJ \, .
\end{eqnarray}
  The integrand has poles at integer $J$ in the left half-plane.
  We explicitly displayed the  rightmost of these poles $1/(J+1)$.  
  It corresponds to $x\sim 1$ integration and gives the leading 
  asymptotic contribution equal to $12 \Lambda^2 /Q^2$.
    Expanding the integrand in the vicinity of $J=-2, -3,$ {\it etc.},
    we can get subleading contributions.
    Note that the singularity at $J=-2$ is a double pole
    $1/(J+2)^2$.  Hence, this contribution will have 
    the $(\Lambda^2/Q^2)^2  \ln(Q^2/\Lambda^2)$ term.  
  One can also close the integration contour in the right half-plane.
  This procedure gives the small-$Q^2$ expansion of 
  $F_{\pi}(Q^2)$,  the first terms of which
   are explicitly written in (\ref{FsmallQ}).


\section{Double distributions in toy model}

\begin{figure}[t]
\mbox{
   \epsfxsize=5.4cm
 \epsfysize=5cm
 \hspace{0cm}
  \epsffile {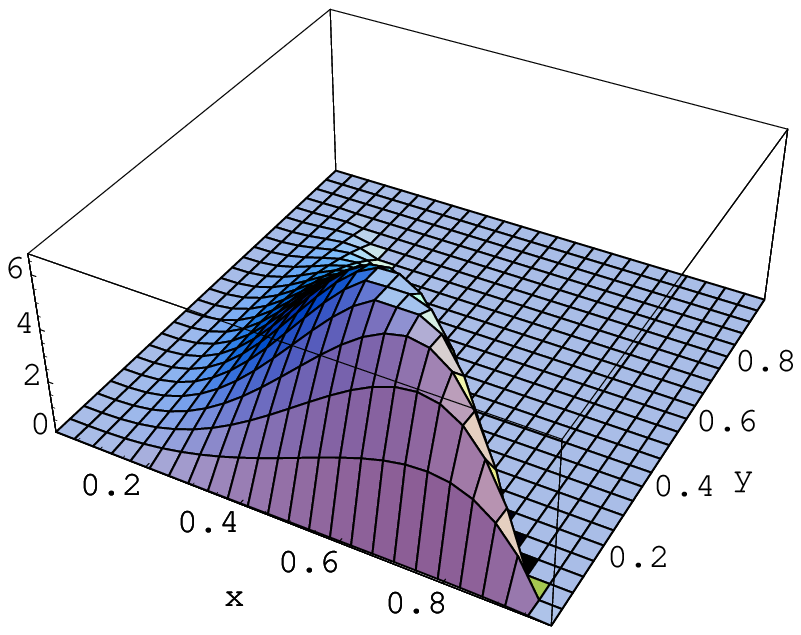} } \hspace{0cm}
\mbox{
   \epsfxsize=5.4cm
 \epsfysize=5cm
 \hspace{0cm}
  \epsffile {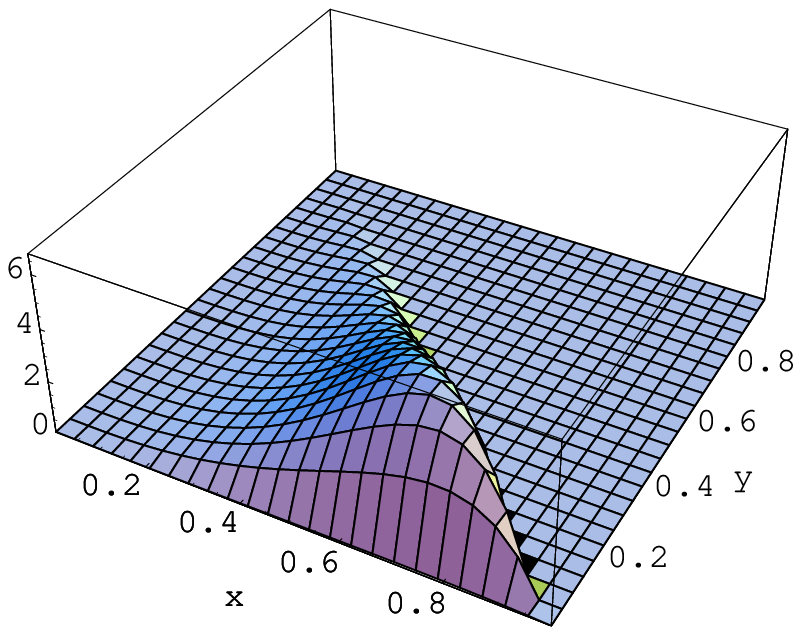} } \hspace{0cm}
\mbox{
   \epsfxsize=5.4cm
 \epsfysize=5cm
 \hspace{0cm}
  \epsffile {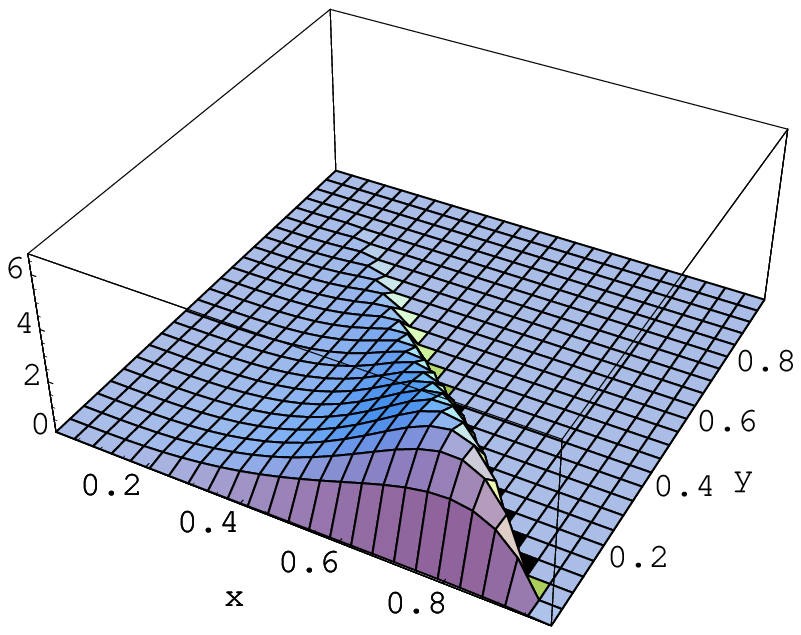} } \hspace{0cm}
\caption{\label{Double-xy}
$F(x,y;t)$ as a  function of $x$ and $y$ for
three values of  $t=0$, $-0.5$ and $-1\,\rm{GeV}^2$.
}
\end{figure}

Let us now analyze the connection of our expression for the pion form factor 
\bea{Fex}
F_{\pi}(Q^2) &=& \frac{1}{A(s)}   \int_0^1 dx 
\int_0^1 d\gamma  \, {6\, \gamma (1- \gamma)  
\over {\left [1+\frac{s^2(x-1/2)^2}{x(1-x)} + 
\gamma (1- \gamma) \frac{s^2 (1-x) Q^2}{4 m^2 x} \right ]^3}}
\end{eqnarray}
with generalized parton distributions.
Introducing the variable $y=(1-x)\, \gamma$, we can rewrite this formula as
\bea{Fybe}
F_{\pi}(Q^2) &=& \frac{1}{A(s)}   \int_0^1 dx 
\int_0^1 dy  \ \theta (x+y \leq 1) \, 
 { 6\, y(1- x-y)  / (1-x)^3\over \left [1+\frac{s^2(x-1/2)^2}{x(1-x)} + 
y (1-x-y)  \frac{s^2  Q^2}{4 m^2 x (1-x)} \right ]^3}\, .
\end{eqnarray}
It  may  be treated as the standard  representation \cite{Radyushkin:1996nd}
 \bea{FDD}
 F_{\pi}(Q^2) = \int_0^1 dx 
\int_0^{1-x}      \, F(x,y;-Q^2) \, dy 
 \end{eqnarray}
 of the pion form factor  in terms of the double distribution\footnote{
 In Appendix A, it is   demonstrated that the variables
 $x$, $y$ in Eq. (\ref{DD}) have the same meaning as 
 in the DD definition.} 
 \bea{DD} 
F(x,y;t) =  \theta (x+y \leq 1) \, { 6\, y(1- x-y) \, 
/ (1-x)^3\over A(s) \, \left [1+\frac{s^2(x-1/2)^2}{x(1-x)} - 
t y (1-x-y)  \frac{s^2  }{4 m^2 x (1-x)} \right ]^3} \, 
  \end{eqnarray}
 (we switched to $t \equiv -Q^2$ to conform with 
  the standard notations for generalized parton distributions).
  This double distribution has correct spectral properties \
  \cite{Radyushkin:1996nd,Radyushkin:1997ki}:
  it vanishes outside the triangle $0 \leq x, y, x+y \leq 1$.
  It also 
  satisfies the Munich symmetry condition \cite{Mankiewicz:1997uy} 
  \bea{Musy} 
F(x,y;t) = F(x,1-x-y;t) \  .
  \end{eqnarray}
 Furthermore, for $t=0$, it has the factorized form  
 suggested in Ref. \cite{Radyushkin:1998es}
 \bea{fact}
  F(x,y;t=0) =   \ \theta (x+y \leq 1) \, h(x,y) \, f(x) \,   ,
 \end{eqnarray}
 in which the $y$-dependence appears only in the 
 normalized profile function
 \bea{profile}
 h(x,y) = 6\, y(1- x-y)  / (1-x)^3
 \end{eqnarray} 
 satisfying 
 \bea{proint} 
 \int_0^{1-x}   \, h(x,y) dy =1 \ .
\end{eqnarray} 
 The remaining factor
 \bea{f(x)} 
 f(x) = {1 \over A(s) \left [1+\frac{s^2(x-1/2)^2}{x(1-x)} \right ]^3} 
 \end{eqnarray}
 depends  on $x$ only, and may be  interpreted as the   
 parton distribution for the valence quarks
inside the pion.
For nonzero $t$, the profile function $h(x,y)$
also factorizes out in the expression for 
the double distribution, Eq.(\ref{DD}). 
However, it is multiplied by a function  
 that  has a  nontrivial
  dependence on  all three variables $x$, $y$ and $t$. 
  In Fig. \ref{Double-xy}, we  plot $F(x,y;t)$ as a  function 
 of $x$ and $y$ for a few values of  $t$.
 
  The form of the double distribution presented above 
  corresponds to parametrization $k = xp_1 +yr$ of the 
  active quark momentum $k$ in terms 
  of  the initial pion momentum $p_1$ and the momentum transfer
  $r=p_1-p_2$.
  The final state pion  has then the momentum $p_2 = p_1-r$:
 the initial and final pions are not treated 
  symmetrically in this formalism.
  To reinforce the symmetry, one should   introduce the  average
  momentum $P=(p_1+p_2)/2$ (the initial and final pion momenta are
 then  $P \pm r/2$, see Section II) and write 
  the 
  active quark momentum as $k =\beta P+ (1+\alpha)r/2$.
  To get the relevant double distribution 
  $f(\beta,\alpha ; t)$, we write the $\gamma$ variable
  as $\gamma = (1+\eta)/2$ and then introduce $\alpha$
  by $\alpha = (1-\beta) \eta$. This gives
    \bea{dd} 
    f(\beta,\alpha ; t) = \theta (| \alpha |  \leq 1-\beta) \, 
     { \frac34 [(1-\beta)^2 - \alpha^2] \, 
/ (1-\beta)^3\over A(s) \, \left [1+\frac{s^2(\beta-1/2)^2}{\beta(1-\beta)} - 
[(1-\beta)^2 - \alpha^2]   \frac{s^2  t}{16 m^2 \beta (1-\beta)} \right ]^3} \, .
     \end{eqnarray}
     The normalized profile function in this case
     is $ \frac34 [(1-\beta)^2 - \alpha^2] \, 
/ (1-\beta)^3$. 
  
  Integration over $k_{\perp}$ can be performed in a similar way for 
   a more general power-law LC wave functions   
   $\psi(x,k_\perp) \sim (a+bk_\perp^2)^{-\kappa}/\sqrt{x(1-x)}$. 
   In this case, 
   one obtains  DDs  with $\kappa$-dependent profiles  
   $\sim [(1-\beta)^2 - \alpha^2]^{\kappa-1}/(1-\beta)^{2 \kappa -1}$.
   The power of the denominator factor in Eq. (\ref{dd}) 
   also changes from 3 to $2 \kappa -1$. 
   Note, that  the faster the decrease of $\psi(x,k_\perp)$ with $k_\perp$,
   the narrower is the $\alpha$-profile of the resulting 
   DD and the faster 
   its decrease with $-t$. The purely exponential 
   wave function ($\kappa \to \infty$) 
   gives an infinitely narrow profile function $\delta (\alpha)$
   (or $\delta [y-(x-1)/2]$ in the case of $F(x,y;t)$). 
   The integral over $\alpha$ (or $y$) is trivial,
   and this is the formal reason why the Gaussian 
   model gives a one-dimensional integral representation for the
   form factor. 
   As suspected, the Gaussian model  is ``too narrow'':  
   it cannot reveal the  $y \ (\alpha)$  profile feature inherent to 
   DDs in general case.

\vskip .2in
\section{Skewed Distributions in Toy Model} 

Having the expression for the double distribution,
 we can construct the nonforward distributions \cite{Radyushkin:1997ki} 
 in the standard way from 
  \bea{NFPD}
  {\cal F}_{\zeta} (X,t) = \int_0^1 dx \int_0^{1-x}  F(x,y;t) \, 
  \delta (X -x -\zeta y) \, dy \ .
   \end{eqnarray}
  Note, that for $F(x,y;t)$ given by Eq.(\ref{DD}),  
  the integrations again can be performed in elementary 
  functions. Hence,   in this particular model,  
  one can analytically  study the interplay
  between $X, \zeta$  and  $t$ dependence of the
  nonforward parton distributions (though the expressions
  are now really  lengthy).

\begin{figure}[t]
\mbox{
   \epsfxsize=5.4cm
 \epsfysize=5cm
 \hspace{0cm}
  \epsffile {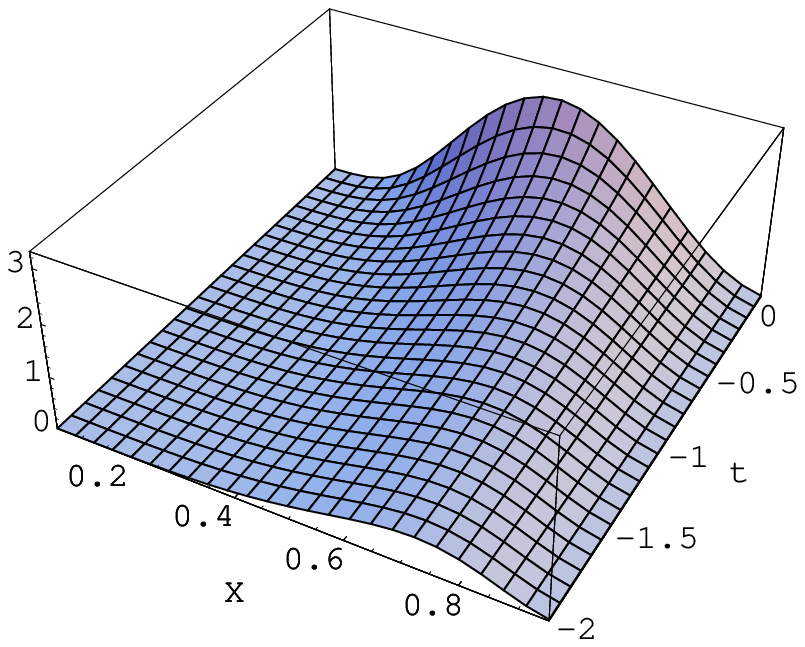} } \hspace{0cm}
\mbox{
   \epsfxsize=5.4cm
 \epsfysize=5cm
 \hspace{0cm}
  \epsffile {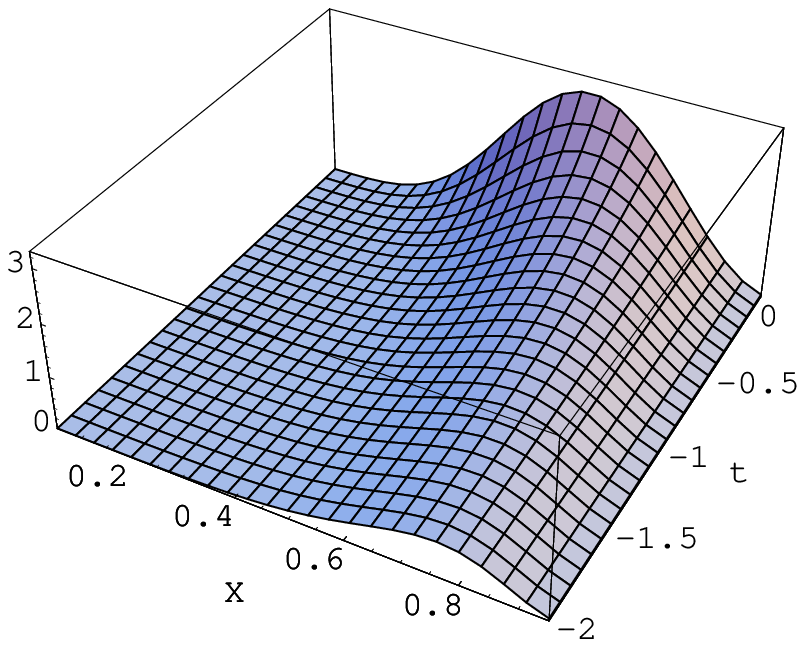} } \hspace{0cm}
\mbox{
   \epsfxsize=5.4cm
 \epsfysize=5cm
 \hspace{0cm}
  \epsffile {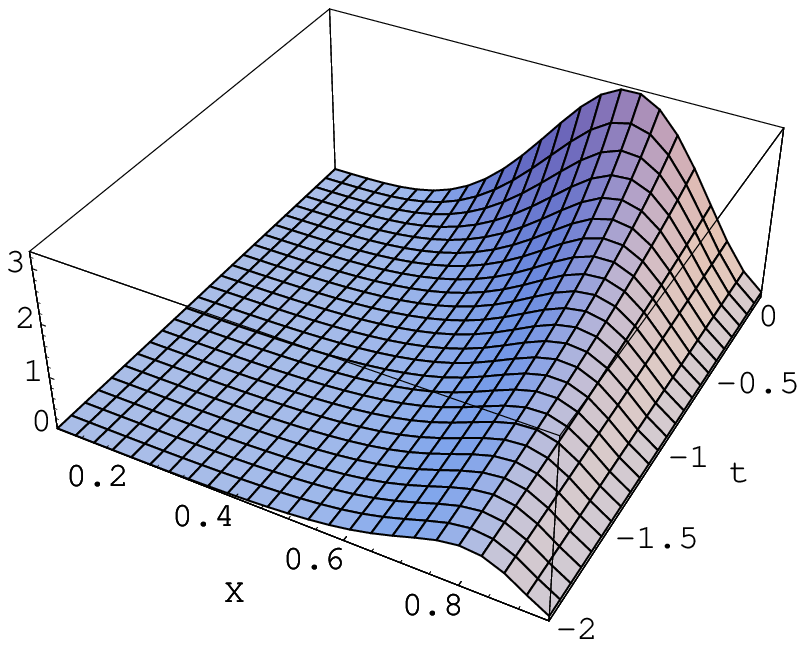} } \hspace{0cm}
\caption{\label{Fzeta-Xt}
${\cal F}_{\zeta} (X,t)$ as a function 
  of $X$ and $t$ for three values of $\zeta=0.2$, $0.4$ and $0.6$.
}
\end{figure}

  In Fig.\ref{Fzeta-Xt},
   we plot ${\cal F}_{\zeta} (X,t)$ as a function 
  of $X$ and $t$ for some values of $\zeta$. 
 Note that   for each value of $\zeta$,  the nonforward distributions
  satisfy the reduction formula
  \bea{redff}
  \int_0^1  {\cal F}_{\zeta} (X,t) \, dX = F_{\pi} (-t) \ . 
   \end{eqnarray}
  An important point is that 
  the right hand side here has no dependence on 
  the skewedness parameter $\zeta$. 
  
 One can also use the symmetric double distribution $f(\beta,\alpha ; t)$ 
 and the relation 
 \bea{OFPD}
 H^{\rm val}(\tilde x,\xi;t) = \int_{0}^1 d\beta \int_{-1+\beta}^{1-\beta} f^{\rm val}(\beta,\alpha;t)  \, 
  \delta (\tilde x-\beta -\xi \alpha) \, d\alpha 
   \end{eqnarray} to 
  obtain Ji's off-forward parton distributions $H^{\rm val}(\tilde x, \xi; t)$.
  Note, that for the infinitely narrow
  profile function \mbox{$f^{\{\infty\} }(\beta,\alpha; t) =F(\beta,t) \delta (\alpha)$}
  corresponding to the purely Gaussian wave function, 
    the OFPD  $H (\tilde x,\xi;t)$ is given by the $\xi$-independent 
  function  $H^{\{\infty\} }(\tilde x,\xi;t) = F(\tilde x,t)$: 
  there are 
    no skewedness effects.  
    Thus, the SPDs obtained from the DDs based on power-law {\it {\it Ansatz}} 
    have a  richer structure.

  \section{``Realistic'' Model} 
  
  The function $f(x)$, Eq.(\ref{f(x)}) was  interpreted 
  above as the toy model version of the 
  valence  quark distribution in the pion. 
  However, its  form
   strongly  differs from the usual  
  phenomenological parametrizations.
  At a normalization point $\mu \sim 1$ GeV,
  the latter  have the form close to 
  $$ f^R (x) =  \frac34 (1-x)/\sqrt{x}\ , $$ 
  with the $1/\sqrt{x}$ reflecting the  Regge behavior
  due to exchanges 
that are not taken into account in the toy   model. 
In the latter,  we assumed that the contribution 
from the higher Fock components to $f(x)$ has the 
same shape as the two-body one.
Also,  the expression that we use
for the two-body wave function
is again just a model guess.
In particular, as shown in  Appendix A,
the double distribution of our toy model
can be obtained from the scalar triangle 
diagram taken at spacelike virtualities  
of the external currents imitating the pions.
In the spin-$1/2$ case, there are 
extra $x$-dependent factors originating 
from the numerators of  propagators.
Thus, one should not take  the 
$x$-dependence of the toy model DD too seriously. 
On the other hand, we observed that the 
$y$-dependence of the model 
DD  has a rather universal structure:
for $t=0$, it is given by the profile
function $\sim [y(1-x-y)]^{\kappa -1}$ only.
For $t\neq 0$, the $y$-dependence 
appears also in the denominator
factor, but it  has a simple structure  basically 
resulting from kinematics. 
These observations suggest 
  to ``minimally correct''   the model DD:
  to change its $x$-shape  
  at $t=0$ without changing the pattern of its 
  $y$ and $t$ dependence. To preserve 
 the   analytic structure 
  of the interplay between the 
   $t$ {\it vs.}  $x$ and $y$ dependence of DDs
   dictated  by the
   simplest {\it {\it Ansatz}} (\ref{DD}) we take the model
   \bea{DDrea} 
   F^{R}(x,y;t) = F(x,y;t) \, \frac{f^R (x)}{f(x)} \  .
      \end{eqnarray} 
 In terms of  the { effective} two-body-like LC wave function,
 this corresponds to the change
  \bea{psirea} 
   \psi(x,k_{\perp}) \to \psi^{R} (x,k_{\perp}) =
   \psi(x,k_{\perp})  \, \sqrt{ \frac{f^R (x)}{f(x}} \  . 
      \end{eqnarray}

\begin{figure}[t]
\mbox{
   \epsfxsize=6cm
 \epsfysize=5cm
 \hspace{1cm}
  \epsffile {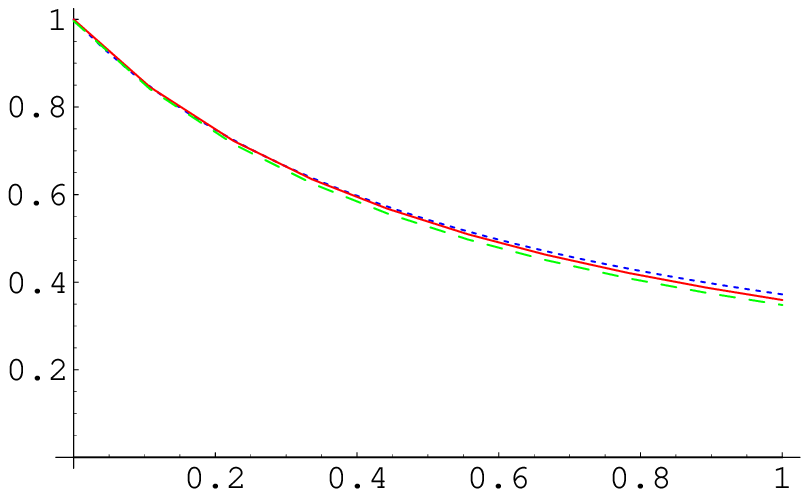} } \hspace{0.5cm}
\mbox{
   \epsfxsize=6cm
 \epsfysize=5cm
 \hspace{1cm}
  \epsffile {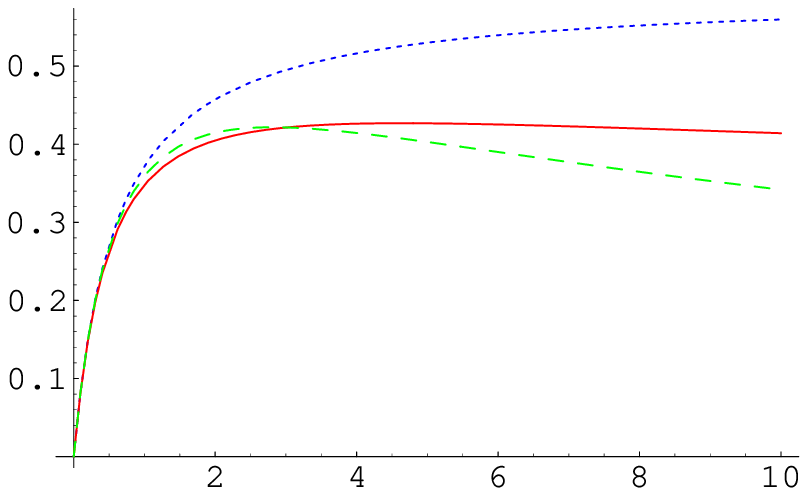} } \hspace{0.5cm}
\caption{\label{FormFactor-new}
Form factor $F_\pi(Q^2)$ (left) and $Q^2 F_\pi(Q^2)\,\rm{GeV}^2$ (right)
obtained with the ``realistic'' model for double distributions (solid line).
For comparison, we present the results for the original model (dashed)
and ``$\rho $ -- meson fit'' $Q^2/(1+Q^2/(0.77\,\rm{GeV})^2)$ (dotted).
}
\end{figure}

    The parameters $m$ and $s$ of the new 
   model should  again be fixed by  
   fitting the slope of the pion form factor
   at $t=0$ and its behavior in the $-t \sim 1\,\rm{GeV}^2$ region.
   In Fig. \ref{FormFactor-new}, we show the curve for the pion form factor 
   obtained with the {\it Ansatz} (\ref{DDrea}) and the values $m=0.46$ and $s=0.81$.
   It practically coincides with the curve obtained within the original model
   for $m=0.3$ GeV and $s=0.95$ in the region of ``not-so-high''
   transfer.
   
\begin{figure}[t]
\mbox{
   \epsfxsize=5.4cm
 \epsfysize=5cm
 \hspace{0cm}
  \epsffile {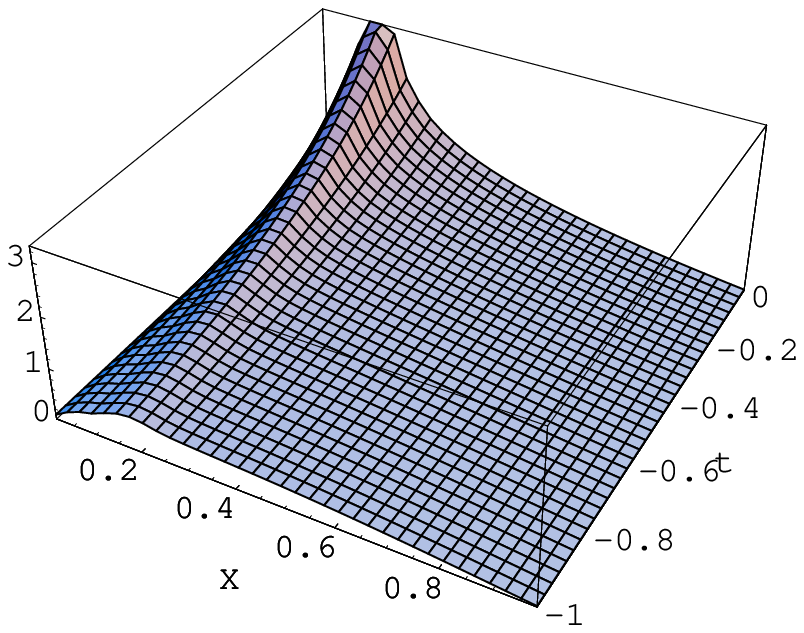} } \hspace{0cm}
\mbox{
   \epsfxsize=5.4cm
 \epsfysize=5cm
 \hspace{0cm}
  \epsffile {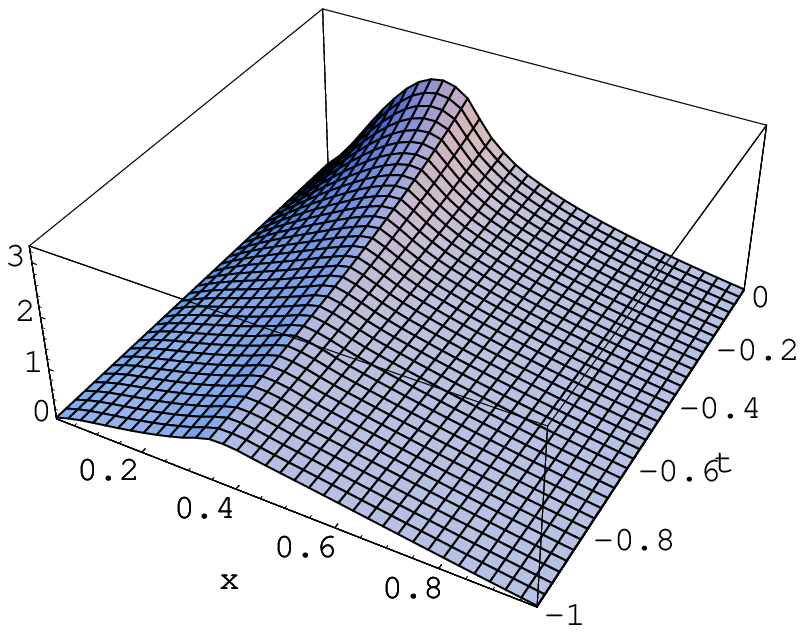} } \hspace{0cm}
\mbox{
   \epsfxsize=5.4cm
 \epsfysize=5cm
 \hspace{0cm}
  \epsffile {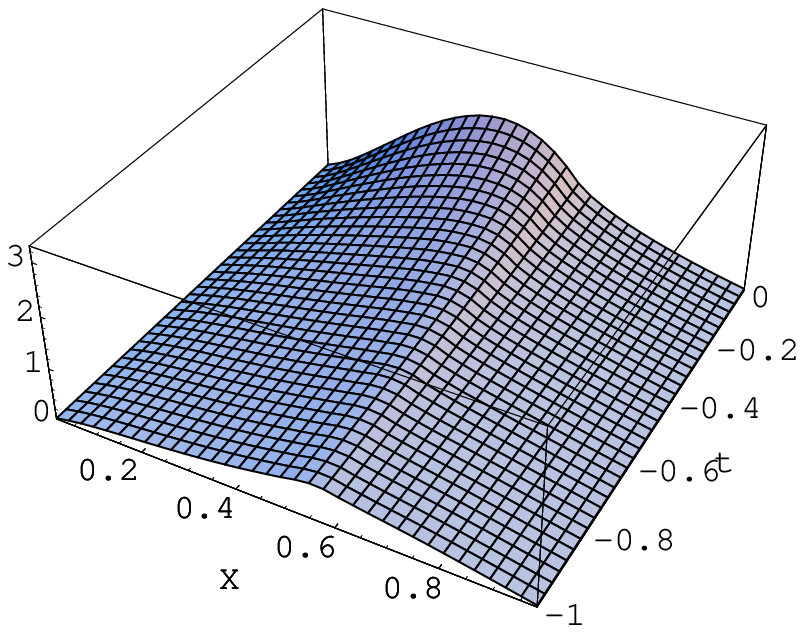} }
\caption{\label{Fzeta-new-Xt}
${\cal F}^R_{\zeta} (X,t)$ obtained with the ``realistic'' DD as
a function of $X$ and $t$ for three values of
$\zeta=0.2$, $0.4$ and $0.6$.
}
\end{figure}

With the new DDs, one can obtain a realistic model for SPDs
${\cal F}^R_\zeta(X,t)$ via (\ref{NFPD}).
The SPD is presented in \mbox{ Fig. \ref{Fzeta-new-Xt}}
 as a function of $X$ and $t$
for some values of $\zeta$.

\begin{figure}[t]
\mbox{
   \epsfxsize=6cm
 \epsfysize=5cm
 \hspace{1cm}
  \epsffile {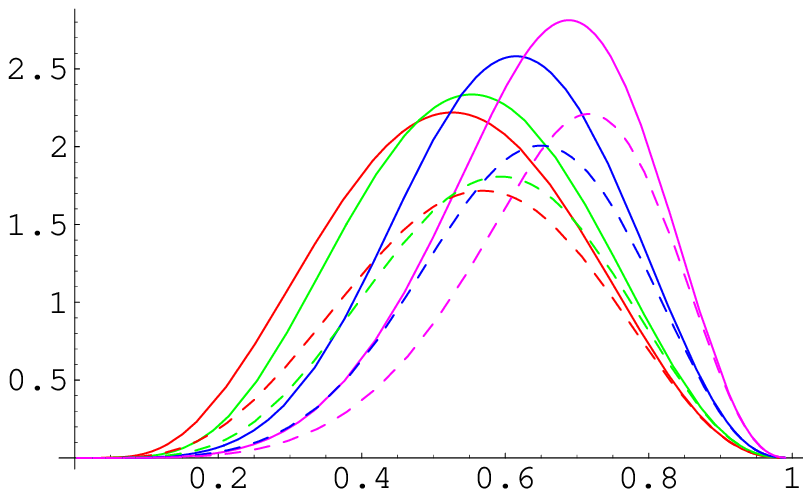} } \hspace{0cm}
\mbox{
   \epsfxsize=6cm
 \epsfysize=5cm
 \hspace{1cm}
  \epsffile {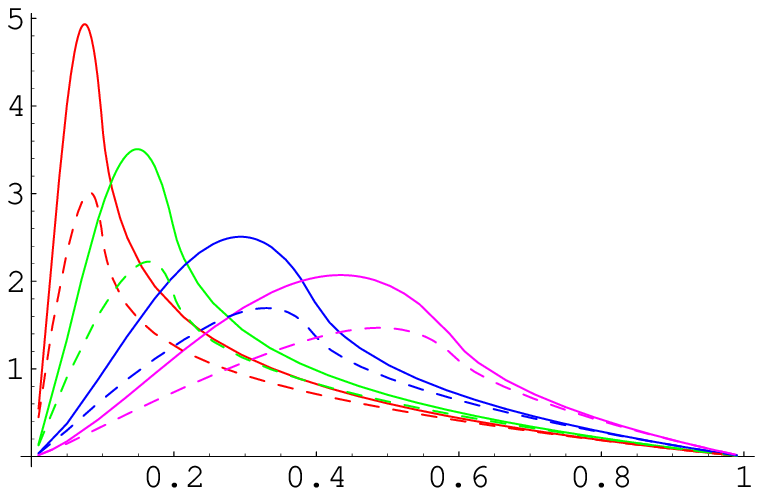} } \hspace{0cm}
  \vspace{3mm}
\caption{\label{SPD-2D.eps}
SPDs $F_\zeta(X;t)$ with $\zeta=0.1$, $0.2$, $0.4$, $0.6$
are shown for two different
values $t=0$ (solid lines) and $t=-0.2\,\rm{GeV}^2$,
both for the original (left) and ``realistic'' (right) model.
The curves corresponding to larger $\zeta$ have maxima at higher $X$.
}
\end{figure}

As a more explicit illustration of the $t$-dependence of  SPDs, in
Fig. \ref{SPD-2D.eps} we show 
SPDs $F_\zeta(X;t)$ with different $\zeta$'s  for two different
values of $t$, both for the original and the ``realistic'' model.

\begin{figure}[t]
\mbox{
   \epsfxsize=6cm
 \epsfysize=5cm
 \hspace{1cm}
  \epsffile {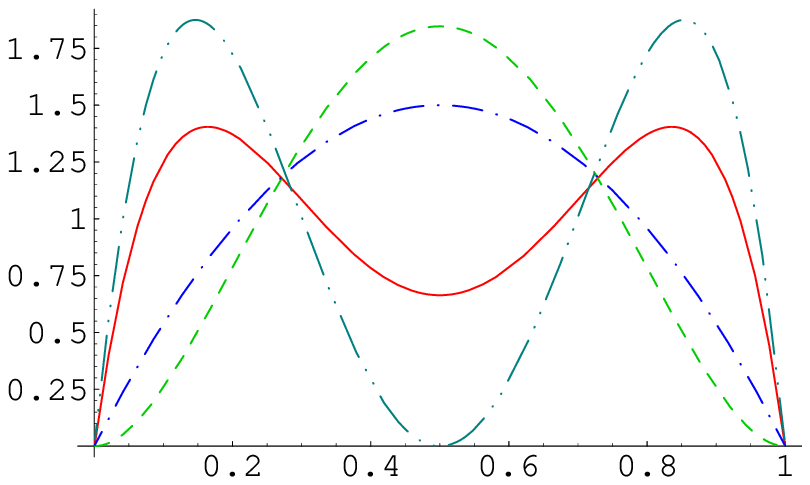} } \hspace{0cm}
\mbox{
   \epsfxsize=6cm
 \epsfysize=5cm
 \hspace{1cm}
  \epsffile {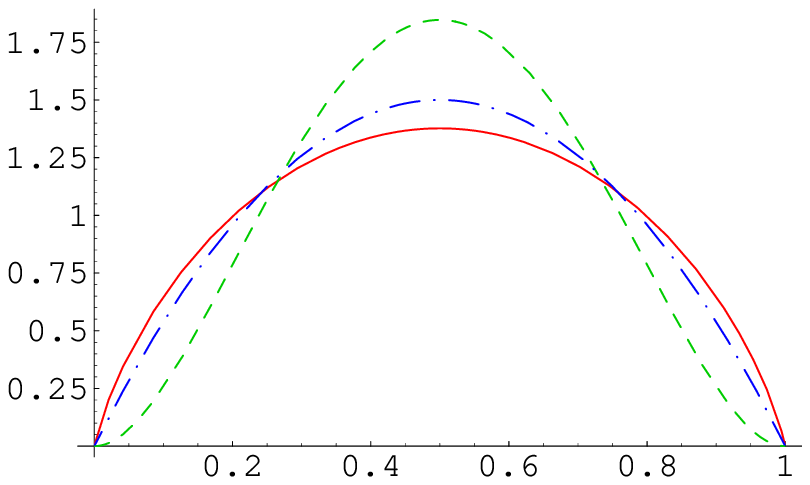} } \hspace{0cm}
  \vspace{3mm}
\caption{\label{pionDAs}
Left: isovector part of  $F_{\zeta=1}(X;t=0)$ 
for the toy model (solid) 
compared to the asymptotic pion DA (dash-dotted),
CZ distribution amplitude (dash-double-dotted) and DA obtained 
from the two-body wave function (dashed). 
Right: isovector part of  $F_{\zeta=1}(X;t=0)$ 
for the realistic  model (solid) 
compared to the asymptotic pion DA (dash-dotted) and 
DA obtained 
from two-body wave function (dashed). 
}
\end{figure}

When $\zeta$ increases, the maxima of SPDs shift to  higher
values of $X$.  The rate of 
change is more drastic in the case of the realistic model,
where 
SPD changes from  a monotonically decreasing 
curve for $\zeta =0$ (corresponding to the usual parton density)
to a shape  resembling that of distribution amplitudes, as $\zeta$ 
tends to 1. It is interesting to analyze  
the limiting case  $\zeta\to 1$. 
As demonstrated  by M. Polyakov 
\cite{Polyakov:1998ze}, in the soft pion limit, $m_{\pi}^2 \to 0, 
\zeta\to 1, 
t=0$,  the isovector part
of the pion SPD should coincide with the pion distribution amplitude.
To check if this constraint is satisfied by our models,
we take the function $F_{\zeta=1}(X;t=0)$ and symmetrize it with respect to 
$X\leftrightarrow 1-X$ to project onto the isovector component. 
In Fig. \ref{pionDAs}, we show the results both for the toy model
 and the realistic model. 
 For the toy model, we obtain a double humped curve,
 resembling the Chernyak-Zhitnitsky (CZ) model $\varphi^{CZ}  (X)=
 30 \, (1-2X)^2 X(1-X)$.
 More precisely, this curve can be fit, with 
 a good accuracy,  by the  sum 
 $0.43 \, \varphi^{as}  (X) + 0.57 \,  \varphi^{CZ}  (X) $ 
 of  the CZ and the asymptotic distribution amplitude  $\varphi^{as}  (X)=
 6  X(1-X)$. 
 The pion distribution amplitude in the toy model can be 
 obtained directly by integrating the two-body wave 
 function $\psi (x, k_{\perp})$ over the transverse momentum. 
 The result is close to the  asymptotic amplitude,
 so one can say that the toy model does not satisfy 
 the constraint imposed by the Polyakov soft pion theorem. 
 On the other hand, in  the realistic model,
 the function $[F_{\zeta=1}(X;t=0) +F_{\zeta=1}(1-X;t=0)]/2$
 is very close to the asymptotic form and to the distribution
 amplitude obtained from the two-body wave function. 
 Experimentally, the pion DA is known to be 
 rather  close to the asymptotic form.
 Thus, the realistic model {\it de facto} 
 satisfies the constraint imposed by the soft pion theorem.
 
\section{Conclusions}    In this paper, we demonstrated 
  how to obtain  a model for the valence 
  (or $C$-odd)  pion double distribution $F (x,y;t)$
  and the  
  skewed parton distribution  ${\cal F}_ \zeta(X;t)$
  satisfying, by construction,  such important constraints 
  as reduction relations to usual parton densities and form factors,
  spectral and polynomiality conditions.   
  SPDs derived in our model have  a nontrivial interplay
  between $X,\zeta$ and $t$ dependence.
  Furthermore, they were adjusted 
  to describe  pion form factor 
  for all available $t$, so we expect that our model 
 describes the $t$-dependence of the 
 pion GPDs of the valence quarks
  both for small and large $t$.
  The ability to have a unified model for  GPDs 
  from $t=0$ to $|t| \sim 10$\,GeV$^2$ is especially
  important in (future) applications to nucleons,
  for which GPDs are already being studied 
  experimentally  both for small (DVCS) and large $t$ 
  (wide-angle Compton scattering). 
 
  \section{Acknowledgements}

The work of I.M. and A.R. was supported by the US 
 Department of Energy  contract
DE-AC05-84ER40150 under which the Southeastern
Universities Research Association (SURA)
operates the Thomas Jefferson Accelerator Facility.
A.R. was also supported by the Alexander
von Humboldt Foundation.
This work has been started 
at the University of Regensburg and continued at the
Ruhr University in Bochum,
and we thank A. Schaefer, V. Braun, K. Goeke, C. Weiss, N. Kivel,
M.V.  Polyakov  and 
P.V. Pobylitsa
for hospitality and fruitful discussions.  
A.R. thanks M. Diehl, G. Miller and B. Tiburzi 
for correspondence, critical comments and 
stimulating discussions.

\begin{appendix}

\section{DD IN A SCALAR MODEL}

Our conversion of the integral representation
 for the form factor 
$F(t)$ into a function $F(x,y;t)$ of three variables 
may look like a rather ambiguous exercize.
Below, by a covariant field-theoretic calculation,
we demonstrate that  $x$ and $y$ 
really have the meaning of the variables of a double distribution.

\begin{figure}[h]
\hspace{4cm}
\mbox{
   \epsfxsize=6cm
 \epsfysize=5cm
 \hspace{0cm}
  \epsffile{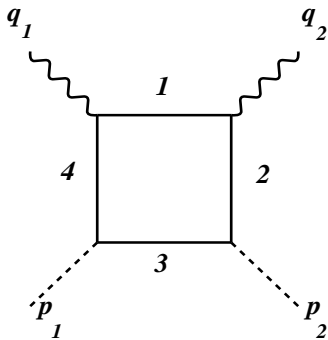}  } \hspace{0cm}
\caption{\label{scaldvcs}
Box diagram for DVCS in a scalar model.
}
\end{figure}

First, consider a one-loop box diagram for the scalar analog of 
deeply virtual 
Compton scattering amplitude (see Fig. \ref{scaldvcs}).  
The  initial and final ``pions'' are
imitated by scalar currents $\Pi$ corresponding to spacelike momenta
$p_1$ and $p_2$, the initial ``photon'' momentum is $q_1$,
and that of the final one is $q_2$.
The momentum invariants describing this 4-point function are
\begin{equation}
p_1^2, p_2^2, Q^2=-q_1^2, q_2^2 =0, s=(p_1+q_1)^2, t=(p_1-p_2)^2 \ . 
\end{equation}
Using the $\alpha$-representation 
\begin{equation}
\frac1{m^2-k^2 - i\epsilon} =
{i} \int_0^{\infty} e^{i\alpha (k^2-m^2+i\epsilon)} 
\, d \alpha
\label{1} \end{equation}
for each of four  scalar propagators 
and calculating  the resulting Gaussian integral over the
loop momentum $k$ we obtain 
\begin{equation}
T(p_1,p_2,q_1) = 
- 
\int_0^{\infty}  \exp\left \{  i \left [{ {
\alpha_1 (\alpha_3 s - \alpha_4 Q^2) + \alpha_2 \alpha_4 t 
+ \alpha_3 (\alpha_4 p_1^2 +\alpha_2 p_2^2)} 
\over{\alpha_1+\alpha_2+\alpha_3+\alpha_4}}
-\rho \, (m^2-i\epsilon)  \right ] \right \}
\frac{d\alpha_1 d\alpha_2 d\alpha_3 d\alpha_4 }{\rho^2}\, , 
\end{equation}
where $\rho \equiv 
\alpha_1+\alpha_2+\alpha_3+\alpha_4$. 
We are interested in the Bjorken kinematics when there are 
two large variables $s$ and $Q^2$ which have the same order of magnitude
$s \sim (1/x_{Bj}-1) Q^2$, while other invariants are small.
The large-$Q^2$ asymptotics in this situation is determined by integration
over the region where the coefficients accompanying $s$ and $Q^2$ 
vanish simultaneously.  Otherwise, the integrand rapidly 
oscillates and the result of  integration 
is exponentially suppressed. Integration 
over $\alpha_1 \sim 0$ region is 
the simplest 
(and, as inspection shows, the leading) possibility. 
It corresponds to hard momentum flow
through the propagator connecting the photon vertices. 
 Performing the $\alpha_1 \sim 0$ integration,  we obtain 
\begin{equation}
T(p_1,p_2,q_1) = 
-i
\int_0^{\infty} 
{\frac{d\alpha_2 d\alpha_3 d\alpha_4/\lambda^2}
{s \alpha_3  /\lambda  -Q^2 \alpha_4/\lambda + i\epsilon }}
\exp \left \{  \frac{i }{\lambda} \left [ \alpha_2 \alpha_4 t 
+ \alpha_3 (\alpha_4 p_1^2 +\alpha_2 p_2^2) \right ] -
i\lambda (m^2-i\epsilon)\right \} + O(1/Q^4) \ ,   
\label{3} \end{equation}
where $\lambda \equiv  \alpha_2+\alpha_3+\alpha_4$.
Denoting $\nu = 2(p_1 q_1)$ and writing $s=\nu -Q^2$
($p_1^2$ is neglected compared to $Q^2$), 
we represent the $Q^2$-dependent term in the 
denominator as \mbox{$\nu \alpha_3 /\lambda - Q^2 (\alpha_3+\alpha_4) /\lambda$},
or finally as  \mbox{$\nu [\alpha_3 /\lambda - \zeta  (1-\alpha_2 / \lambda)]$},
where $\zeta = Q^2/\nu$. 
Introducing  the double distribution 
\begin{equation}
F(x,y;t,p_1^2,p_2^2) = i 
\int_0^{\infty} \, 
\delta \left ( x - \frac{\alpha_3}{\lambda} \right   ) 
\delta \left ( y - \frac{\alpha_2}{\lambda}
\right   ) \, 
\exp \left \{  \frac{i }{\lambda} \left [ \alpha_2 \alpha_4 t 
+ \alpha_3 (\alpha_4 p_1^2 +\alpha_2 p_2^2) \right ] -
i\lambda (m^2-i\epsilon)\right \} 
 \frac{d\alpha_2 d\alpha_3 d\alpha_4}{\lambda^2} \, ,
\label{16} \end{equation}
we can write the scalar DVCS amplitude in the partonic form
\begin{equation}
T(p_1,p_2,q_1) = 
-\frac1{\nu} \int_0^1 \int_0^1 
\frac{F(x,y;t,p_1^2,p_2^2)}{x+y\zeta -\zeta+i\epsilon} 
\,    dx \, dy  \   .
\end{equation}

A few comments are in order.
First, we reemphasize  that    
there is no $\zeta$-dependence in the definition of DDs.
The second comment concerns the spectral 
properties of DDs. It  is easy to see that both  variables $x,y$ vary
between 0 and 1. Furthermore, their sum is also  
confined within these limits:
$0 \leq x+y \leq 1$. 
Finally, the hard amplitude depends on the 
DD variables $x,y$ through the combination 
$x+y\zeta$ only, so one can treat it as a new variable
$X=x+y\zeta$ and  use nonforward  parton distributions ${\cal F}_{\zeta} (X;t)$
instead of  DDs $F(x,y;t)$. 

With the same  technique, one can 
calculate the nonforward matrix element 
$\langle p_2| {\cal O}_n |p_1 \rangle$  
of the composite operator ${\cal O}_n= \varphi (iz \vec \partial)^n\varphi$ 
and obtain its $\alpha$-representation
 \begin{equation}
i 
\int_0^{\infty} \, 
 \left ( \frac{\alpha_3 (p_1z) + \alpha_2 (rz)}{\lambda} \right   )^n 
\exp \left \{  \frac{i }{\lambda} \left [ \alpha_2 \alpha_4 t 
+ \alpha_3 (\alpha_4 p_1^2 +\alpha_2 p_2^2) \right ] -
i\lambda (m^2-i\epsilon)\right \} 
 \frac{d\alpha_2 d\alpha_3 d\alpha_4}{\lambda^2} \, .
\label{16a} 
\end{equation}
Note, that that the derivative $(iz \vec \partial)$ acting on field $\varphi$ 
is expected to give the factor $(kz)$,
where $k$ is the momentum of $\varphi$. Eq.(\ref{16a})
shows  that $(kz) = \alpha_3 (p_1z)/\lambda + \alpha_2 (rz)/\lambda$,
{\it i.e.,} $\alpha_3 /\lambda$ and $\alpha_2 /\lambda$ 
should be interpreted as the variables $x$ and $y$
of the double distribution $F(x,y;t)$. 
Alternatively, using the binomial expansion for  the $( \ldots)^n$ 
factor, one can  see that  the coefficients in front 
of $(p_1 z)^{n-k} (rz)^k$ in this expansion are  given
by the $x^{n-k} y^k$ moments of the DD (\ref{16}). 
Integrating over $\lambda$, we get   the relevant DD  explicitly
\begin{equation}
F(x,y;t,p_1^2,p_2^2)= - 
\left \{y (1-x-y) t + x [(1-x-y)p_1^2 +yp_2^2]-m^2 \right \}^{-1}\ . \label{FV}  
\end{equation}
Putting the ``pions'' on equal footing by setting 
$p_1^2 = p_2^2 =-M^2$, we get a DD
\begin{equation}
F^{(1)} (x,y;t,M^2)=  \left \{-y (1-x-y) t + x (1-x)M^2 +m^2 \right \}^{-1}
\end{equation}
satisfying the  $y \leftrightarrow 1-x-y$ {\it Munich
symmetry}  condition. 
This expression can be rewritten in the form 
\begin{equation}
F^{(1)}(x,y;t,M^2)= \frac1{x(1-x)M^2} 
\left \{1+\frac{m^2} {x (1-x)M^2}  -\frac{y (1-x-y) t}{x (1-x)M^2} 
 \right \}^{-1}
\end{equation}
resembling 
the DDs obtained from the power-law wave function.
Introducing  $\Lambda^2 = M^2/4 +m^2$, we get the expression
\begin{equation}
F^{(1)}(x,y;t,M^2)|_{M^2=4 (\Lambda^2-m^2)}= \frac1{4 x(1-x)\Lambda^2} 
\left \{1+\frac{m^2(x-1/2)^2} {4x (1-x)\Lambda^2}  -
\frac{y (1-x-y) t}{4x (1-x)\Lambda^2} 
 \right \}^{-1}
\end{equation}
whose denominator factor has the structure  close  
to that of the DDs obtained in the model with the power-law wave functions.
However, the denominator power is $(-1)$ instead of $(-3)$. 
Applying $(p_1^2 \partial / \partial p_1^2) (p_2^2 \partial / \partial p_2^2)$ 
to $F(x,y;t,p_1^2,p_2^2)$, Eq. (\ref{FV})\footnote{It is easy to see   that
the $\Pi \varphi \varphi$ scalar vertex differentiated
with respect to the virtuality of the scalar $\Pi$-current 
corresponds to the $\kappa =2$ power-law wave function
(\ref{psi}).},    
and setting $p_1^2 = p_2^2 =-M^2$, we obtain 
\begin{equation}
F^{(2)}(x,y;t,M^2)= \frac{2 y (1-x-y)}{x(1-x)^3 M^2} 
\left \{1+\frac{m^2}{x (1-x)M^2}  -\frac{y (1-x-y) t}{x (1-x)M^2} 
 \right \}^{-3} \  .
\end{equation}
 Now, using $\Lambda^2 = M^2/4 +m^2$, we end up with 
the DD differing from the toy model DD, Eq. (\ref{DD}) just 
by the $x$-dependent factor $1/x$ and an overall normalization.
Note, that  our toy DD was based 
on the formula for the vector  form factor,
 hence, for full correspondence, we  should consider DD related  
to operators 
containing the  extra  $i\stackrel{\leftrightarrow}{\partial^{\mu}}$
derivative.
 This  results in the extra factor of 
 $$\frac{\alpha_3}{\lambda} \left (p_1^{\mu} + p_2^{\mu} \right )
 + \frac{\alpha_2  - \alpha_4}{\lambda} \left (p_1^{\mu} - p_2^{\mu} \right )
 =2xP^{\mu}+(2y-x-1)r^{\mu}  \ . $$ 
 As expected, the $P^{\mu}$-part 
 contains the missing factor of $x$. 
 Since the $r^{\mu}$-part is {\it Munich-antisymmetric}, 
 it does not contribute to form factor 
and forward densities (see also \cite{Tiburzi:2002sxyz}). 
In general, such terms    are not  restricted
by the reduction relations (\ref{redf}), (\ref{redF}). 
However, they  contribute to SPDs for $\zeta \neq 0$,
and their modeling deserves  a separate consideration.

\section{SPDs IN THE IMPACT PARAMETER REPRESENTATION}

 In this Appendix, we investigate  the 
properties of our model SPDs in the impact parameter representation.
For the scalar triangle diagram, such an analysis 
was recently performed by Pobylitsa \cite{Pobylitsa:2002vw}.
He also uses the $\alpha$-representation and double distributions 
for the triangle diagram, but   
takes the version of double distributions in which 
the plus component of the  momentum of the spectator system 
is written as $up_1^+ + vp_2^+$. Instead of 
Eq.(\ref{16}), we would have  then
\begin{equation}
P(u,v;t,p_1^2,p_2^2) = i 
\int_0^{\infty} \, 
\delta \left ( u - \frac{\alpha_4}{\lambda} \right   ) 
\delta \left ( v - \frac{\alpha_2}{\lambda}
\right   ) \, 
\exp \left \{  \frac{i }{\lambda} \left [ \alpha_2 \alpha_4 t 
+ \alpha_3 (\alpha_4 p_1^2 +\alpha_2 p_2^2) \right ] -
i\lambda (m^2-i\epsilon)\right \} 
 \frac{d\alpha_2 d\alpha_3 d\alpha_4}{\lambda^2} \, .
\label{16uv} \end{equation}
This DD is related to SPD $H(\tilde x,\xi;t)$ by
 \begin{equation}
H(\tilde x,\xi;t, p_1^2,p_2^2) = \int_0^1 \int_0^1 
 \delta (1- \tilde x -u(1+\xi) - v (1-\xi)) \, 
P(u,v;t,p_1^2,p_2^2) \, \theta (0\leq u+v \leq 1)\, du\, dv \  .
\label{Huv} 
\end{equation}
Now,  one should take   $p_1^2=p_2^2 = m_\pi^2$,
 $t=-(|\Delta_{\perp}|^2+4\xi^2 m_\pi^2)/(1-\xi^2)$
and calculate the double Fourier transform
\begin{equation}
B(\tilde x,\xi;b_{\perp}) = \int \frac{d^2 \Delta_{\perp}}{(2\pi)^2}\, 
e^{i(\Delta_{\perp} b_{\perp})} H \left (\tilde x,\xi; - 
\frac{|\Delta_{\perp}|^2+4\xi^2 m_\pi^2}{1-\xi^2 } \right)  \ .
\label{Bb} 
\end{equation}
The  $\delta$-function in Eq.(\ref{Huv})
can be rewritten as $\delta (1-u/r_1 -v/r_2)/(1-\tilde x)$,
where the parameters $r_{1}$,  $r_{2}$ given by
$r_1=(1-\tilde x)/(1+\xi)$, $r_2 =(1-\tilde x)/(1-\xi)$ have the 
meaning of the spectator's plus-momentum measured 
in units of the initial or final pion plus-momenta.
Due to this $\delta$-function, we can write $u=zr_1,  v=(1-z)r_2$,
with $0\leq z \leq 1$ in the $\tilde x >\xi$ region. 
Finally, the integral over $d\lambda dz$ can  be transformed 
into  integration  over 
the variables $\sigma_1=z\lambda$ and $\sigma_2=(1-z) \lambda$.
A remarkable fact is that the resulting integrand $I(\sigma_1,\sigma_2)$ 
factorizes $I(\sigma_1,\sigma_2) = J_1 (\sigma_1) J_2(\sigma_2)$. 
As a consequence, the expression for  $B(\tilde x,\xi;b_{\perp})$ 
also has a factorized form \cite{Pobylitsa:2002vw}
\begin{equation}
B(\tilde x,\xi;b_{\perp}) = \frac{1-x}{4 \pi} \,
V_0(r_1,(1-\xi) b_{\perp}) \, V_0(r_2,(1+\xi) b_{\perp}) \  ,
\label{Bbf} 
\end{equation}
where the generalized impact-parameter LC wave functions 
\begin{eqnarray}
V_0(r, c_{\perp}) = \frac1{4\pi r}\int_0^{\infty} \frac{d\sigma}{\sigma} 
\exp \left [-\frac{c_{\perp}^2}{4\sigma r^2}
-\sigma (m^2 -r(1-r)m_\pi^2) \right ] \nonumber \\ =
\frac1{2\pi r} K_0\left (\frac{|c_{\perp}|}{r}
\sqrt{m^2 - r(1-r)m_\pi^2)} \right ) 
\end{eqnarray}
can be expressed through   the modified Bessel function
$K_0$.  As demonstrated in Refs. \cite{Pobylitsa:2002vw}, the factorized representation 
(\ref{Bbf}) guarantees that the positivity bounds 
\cite{Martin:1997wy,Radyushkin:1998es,Pire:1998nw,Ji:1998pc,Diehl:2000xz,Pobylitsa:2001nt,Pobylitsa:2002gw,Diehl:2002he}
for this 
SPD are satisfied in the  model with scalar quarks.
The same SPD multiplied by $(1-\tilde x)$ satisfies 
the positivity bounds for spin-$1/2$ quarks.

If we would proceed as in   Appendix  A,
$i.e.,$ first  differentiate Eq.(\ref{16uv}) with respect to
$p_1^2$ and $p_2^2$ and  then take $p_1^2= p_2^2 = m_\pi^2$,
we would  get an  extra factor $\alpha_2 \alpha_4 \alpha_3^2/\lambda^2$. 
After the transformations described above, this would result 
in the factor $ r_1 r_2 \sigma_1\sigma_2 
[1-(r_1 \sigma_1 + r_2 \sigma_2)/(\sigma_1+\sigma_2)]^2$,
and the integral over $ \sigma_1, \sigma_2$ cannot be factorized 
into a product of two separate integrals over $\sigma_1,\sigma_2$.
The unfactorizable piece comes from the $\alpha_3^2$ factor
resulting   from differentiation with respect to external 
virtualities. To avoid this factor, but still preserve 
the $\sim 1/k_{\perp}^4$ behavior of the effective IMF wave function,
one can perform differentiation with respect to the squares of the 
quark masses  ($i.e.,$ take all the  masses different, 
differentiate   with respect to $m_i^2$'s corresponding 
to lines 2 and 4 and then take all the masses equal).
This would produce the factor $\alpha_2 \alpha_4$, or eventually
$ r_1 r_2 \sigma_1\sigma_2$, which does not violate 
the  factorized structure of the integrand. In the
impact parameter representation, the  result 
has the structure  of Eq.(\ref{Bbf}), but with $V_0(r,c_{\perp})$ 
substituted by the expression 
\begin{eqnarray}
V_1(r, c_{\perp}) = \frac{|c_{\perp}| m}{4\pi \sqrt{1- r(1-r)m_\pi^2/m^2}}\, 
 K_1\left (\frac{|c_{\perp}|m}{r}
\sqrt{1- r(1-r)m_\pi^2/m^2 } \right ) 
\end{eqnarray}
involving the modified 
Bessel function
$K_1$.  For the original IMF  wave function, differentiation
with respect to  the active quark mass is equivalent to choosing
the {\it Ansatz} 
\bea{psi2}
\psi(x,k_\perp)= \frac{\cal N}{x\sqrt{x(1-x)}[a+bk_\perp^2]^2}
\end{eqnarray}
instead of Eq.(\ref{psi}).
It has the extra $1/x$ factor enhancing the wave function 
at small $x$. In the spirit of our discussion 
of the $x$-dependence in the main text of the paper,   
we may say that such a function more adequately models 
 the contribution of higher Fock components.
  
The  positivity bounds are satisfied also in a more general 
case when $B(x,\xi;b_{\perp})$ is given by a sum \cite{Pobylitsa:2002vi,Pobylitsa:2002vw} 
\bea{Bgen}
B(\tilde x,\xi;b_{\perp}) = (1-\tilde x)^{N+1} \sum_{n} Q_n(r_1,(1+\xi)b_{\perp})
Q_n^*(r_2,(1-\xi)b_{\perp})  \ ,
\end{eqnarray}
where $N=0$ for ``scalar quarks'' and $N=1$ for spin-$1/2$ case.
This opens a possibility to build models for GPDs
consistent  both with the polynomiality and 
 positivity constraints. 
The simplest idea is to start with 
the $\alpha$-representation (\ref{16}) for DD
corresponding to the scalar triangle diagram, and modify it
by multiplying the integrand by a function $R(m^2\sqrt{\alpha_2 \alpha_4})$ 
 depending only
on  the product $\alpha_2 \alpha_4$ (choosing the 
argument as $\sqrt{\alpha_2 \alpha_4}$ we get eventually a function of $(1-\tilde x)$;
the parameter $m^2$ was included to make the argument of $R(a)$ dimensionless). 
 If  this function 
can be represented as 
\bea{R}
R(a)=\sum_n R_n a^n
\end{eqnarray}
with all $R_n$ positive (the sum should be understood in a 
wide sense, it can involve integrations), 
the  function $B(\tilde x,\xi;b_{\perp})$
in such a model would have the structure of Eq.(\ref{Bgen}),
and positivity constraints are satisfied. 
The model double distribution 
based on  $R(a)$ gives the following expression for $H(\tilde x,\xi;t)$
in the $\tilde x > \xi$ region 
 \bea{HR}
H(\tilde x,\xi;t)|_{\tilde x > \xi} 
= (1-\tilde x)^{N-1} r_1 r_2 \int_0^{\infty} d\rho \,
\int_0^1 dz \, R(\rho \sqrt{z \bar z \,   r_1 r_2} )\nonumber \\ 
\times \exp \left \{ - \rho \, [1- (1-zr_1-\bar z r_2)(zr_1+
\bar z r_2)m_\pi^2/m^2 -z\bar z\, r_1 r_2 t/m^2 ] \right \} 
\end{eqnarray}
(we   use  the notation 
 $\bar z \equiv  1- z$ here, and later we  also use $\bar x \equiv  1- x$,
 {\it etc.}).   Note  that 
 we performed Wick rotation
$\alpha_j  \to -i \alpha_j$  in the original
$\alpha$-representation (\ref{16uv}),  which is justified 
if the pion stability condition $m_\pi^2 < 4m^2$ is satisfied. 
In the forward limit ($\xi=0, t=0$) this gives 
\bea{Mfor}
f(x) = (1-x)^{N+1} \int_0^1 d z \int_0^{\infty} d\rho \,
R(\rho \sqrt{z \bar z} \, \bar x ) \, 
e^{-\rho (1  - x\bar x m_\pi^2/m^2)} \ .
\end{eqnarray}
The function $R(a)$ should be    adjusted to
fit  experimental forward distribution $f^{\rm exp}(x)$,
and then it can be used for calculation 
of $H(\tilde x,\xi;t)$. In particular,  for massless pion,
the coefficients $R_n$ can be expressed directly 
\bea{Rn}
R_n= \frac{n+1}{[\Gamma (n/2+1)]^2}  \, A_{n+1+N}
\end{eqnarray}
in terms of the coefficients of the $\bar x^k$ expansion 
of the forward distribution 
\bea{fk}
f^{\rm exp}(x) = \sum_k A_k (1-x)^k \ .
\end{eqnarray}
Note, that for the simple model  $f^R(x)=\frac34 (1-x)/\sqrt{x}$
of
 Section IX, all the coefficients
$A_k=\Gamma (k-1/2)/ \Gamma (1/2) (k-1)!$ are positive.
Alternatively, since both Eqs.(\ref{HR}) and (\ref{Mfor})
in the  $t=0, m_\pi^2=0$ limit involve the same integral
of the $R$-function,  the only change being  $\bar x \to \sqrt{r_1r_2}$, 
in this case we can  directly write $H$ through $f(x)$ 
\bea{Hf}
H(\tilde x,\xi;t=0)|_{\tilde x > \xi \, ; \, m_\pi^2=0} = 
\left (\frac{1- \tilde x}{\sqrt{r_1 r_2}} \right )^{N-1}
f(1-\sqrt{r_1 r_2})=(1-\xi^2)^{(N-1)/2}f\left (1-\frac{1-\tilde x}{\sqrt{1-\xi^2}}\right ) \ .
\end{eqnarray}
In the case of nonforward distributions,
we have 
\bea{Ff}
{\cal F}(X,\zeta;t=0)|_{X > \zeta \, ; \, m_\pi^2=0} = (1-\zeta)^{(N-1)/2}
f\left (1-\frac{1-X}{\sqrt{1-\zeta}}\right ) \ .
\end{eqnarray}
Taking $f(x) = f^R(x)$ and $N=1$ (spinor quarks),
we get the curves which  are very close to 
the $X>\zeta $ parts of the realistic model curves 
shown in Fig. 7. 
In the region $X\leq \zeta$, the functions can be obtained 
only from formulas explicitly involving  double  distributions. 
In particular, the  scalar $R(a)$ model gives
\bea{RaMDD}
F(x,y;t) = \int_0^{\infty} \, e^{-\rho \, p(x,y,t) } \,   
R[\rho \, \sqrt{y(1-x-y)} ] \, d\rho \ ,
\end{eqnarray}
where $p(x,y,t) = 1-y(1-x-y)t/m^2 - 
x(1-x)m_{\pi}^2/m^2$. Fixing $R(a)$,  {\it  e.g.,} 
 by the requirement  that $f(x) =f^{R}(x)$ 
in the $m_{\pi}\to 0$ limit (which allows to get the result in  analytic form) 
and  using Eq.(\ref{Rn})
we get 
 \bea{RaMDDExp}
 F(x,y;t) = \sum_{n} A_{n+1} \frac {(n+1)![y(1-x-y)]^{n/2}}
 { \Gamma^2 (n/2+1)[p(x,y,t)  ]^{n+1}}  \ .
\end{eqnarray}
 This expression  can be also written as 
\bea{RaMDDProf}
 F(x,y;t) = \sum_{n} A_{n+1} 
 \, h^{(n/2)} (x,y) \left[\frac{(1-x)}{p(x,y,t)} \right ]^{n+1}
\end{eqnarray}
where  $h^{(n/2)} (x,y)$ is 
the  normalized profile function of order $n/2$. 
Thus,  DD $F(x,y;t)$ in this model is given by  a sum of powerlike 
terms similar to those  discussed in Section VII.
The main difference is that the $y$-profile 
now is not universal: one should expand the forward 
distribution $f(x)$ into a  power series over $(1-x)$ and 
supplement  the  $(1-x)^{n+1}$  term by 
 the $n$-dependent profile function
$h^{(n/2)} (x,y)$.   
   Because of the correlation between the power of $(1-x)$ and the order of the
profile function, the shape of the $y$-profile of the double distribution
changes with $x$.  
Since  all  parton  distributions $f(x)$ tend to 
infinity  as $x$ goes to  $0$, the small-$x$ region is dominated 
by terms with large $n$, and the profile
is more narrow when $x \to 0$,  becoming infinitely narrow
as $x\to 0$.  Note, that in the case when the profile is infinitely narrow
for all $x$, {\it i.e.,}  when $ F(x,y;t=0) = \delta [y-(1-x)/2] f(x)$,
the $t=0$ skewed distribution is given by 
${\cal F}_{\zeta}(X) = f[(X-\zeta/2)/(1-\zeta/2)]/\sqrt{1-\zeta} $: 
it repeats the 
form of the forward distibution $f(x)$ and is infinite for $X=\zeta/2$.
To check if this  also  happens for the model (\ref{RaMDDProf}) 
with the changing  profile
and realistic $\sim x^{-0.5}$ behavior for small $x$, 
we 
took $f^{\rm exp}(x) = f^R (x)$ (or, what is the same,
$A_{n+1} = \Gamma (n+1/2)/\Gamma (1/2) \, n!$ in Eq.(\ref{RaMDDExp})) and  
 constructed   
model skewed distributions ${\cal F}_{\zeta}(X)$  both in 
$X>\zeta$ and $X<\zeta$ regions (see Fig.10). 
The resulting functions are indeed singular for $X=\zeta/2$, but 
finite otherwise.

\begin{figure}[t]
\begin{center}
\mbox{
   \epsfxsize=8cm
 \epsfysize=5cm
 \hspace{1cm}
  \epsffile {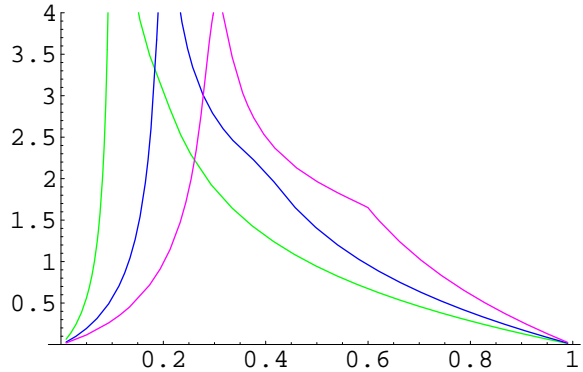} } 
  \end{center}
  \vspace{3mm}
\caption{\label{new.eps}
SPDs $F_\zeta(X;t=0)$ with $\zeta=0.2$, $0.4$, $0.6$
obtained from Eqs.(\ref{RaMDDProf}) and (\ref{fk}).
 The forward distribution was modeled by  $f^{R}(x)=\frac34 (1-x)/\sqrt{x}$.
The curves   tend to $\infty$ for $X=\zeta/2$.
}
\end{figure}

We plan to perform a  more detailed study of GPDs
within the $R(a)$-model in a separate paper.

\section{TWO-PION DISTRIBUTION AMPLITUDE IN SCALAR MODEL}

Another referee's request was to apply our approach 
to processes of two-pion production in $\gamma^* \gamma \to \pi \pi$
or $\gamma \gamma \to \pi \pi$ collisions. 
The $\gamma^* \gamma \to \pi \pi$
process  in the kinematics when $\gamma^*$ is highly virtual,
while $s \equiv m_{\pi \pi}^2$ is small \cite{Diehl:1998dk} is  the  crossed channel
 reaction to 
DVCS while  
$\gamma \gamma \to \pi \pi$ process in the kinematics when $s, |t|,|u|$ are large
\cite{Diehl:2001fv} is    the crossed channel reaction 
to wide-angle Compton scattering.
Here we are going to  consider only the simpler case of the 
kinematics of the first process.

\begin{figure}[h]
\hspace{4cm}
\mbox{
   \epsfxsize=9cm
 \epsfysize=5cm
 \hspace{0cm}
  \epsffile{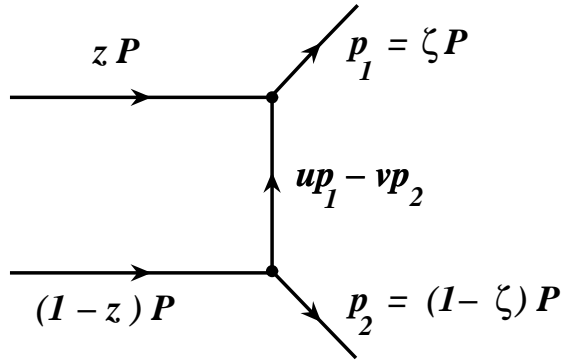}  } \hspace{0cm}
\caption{\label{2piDA}
Plus-momentum flux  structure of the two-pion distribution amplitude.
}
\end{figure}

Originally \cite{Diehl:1998dk}, it was proposed to describe 
the nonperturbative stage of this process by  two-pion distribution amplitude (2$\pi$DA) 
$\Phi(z,\zeta;s)$ which describes the conversion 
of two quarks with plus-momenta $zP^+$ and $\bar z P^+$ 
into two pions with momenta $p_1^+ = \zeta P^+$ and $p_2^+ = \bar \zeta P^+$ 
(see Fig. \ref{2piDA}), where $P$ is the total momentum of the pion pair 
(recall that the invariant mass of the pair is small: $P^2 \equiv s\ll 1\,$GeV$^2$). 
Later,  Teryaev \cite{Teryaev:2001qm} proposed to use the double distribution  description 
(see also Ref. \cite{Kivel:2002ia} for further developments) corresponding 
to parametrization of  the plus-component of the spectator momentum as 
$up_1^+ - vp_2^+$ (note, that both momenta are now outgoing, and 
this is reflected in  the change of the relative sign of $p_1$ and $p_2$
parts of the spectator momentum  compared to the DVCS case).  
In $P^+$ units, the spectator plus-momentum 
can be written  either as 
$(\zeta-z)P^+$ (using the 2$\pi$DA variable $z$)
or as $u\zeta P^+ -v \bar \zeta P^+$ (using the DD variables $u,v$).
Thus, the connection between the two sets of variables
is given by $$z=(1-u-v)\zeta +v\ .$$
So, we can express the 2$\pi$DA 
$\Phi(z,\zeta;s)$ in terms of the DD $M(u,v;s)$:
\bea{Pzzetas}
\Phi(z,\zeta;s)= \int_0^1 du \int_0^1 dv \, \theta (u+v  \leq 1)\, 
\delta (z-\zeta (1-u-v) - v) \, M(u,v;s)
\ .
\end{eqnarray}
The DD  representation for $\Phi(z,\zeta;s)$ can be used to derive
some general properties of 2$\pi$DAs, such as  
the polynomiality  condition
\bea{polyn}
\int_0^1 z^n \, \Phi(z,\zeta;s)\, dz = \sum_{l=0}^n \, K_l \, \zeta^l
\ ,
\end{eqnarray}
which  states that  $z^n$ moment of $\Phi(z,\zeta;s)$ is $n$th order  polynomial of $\zeta$.

To model  2$\pi$DAs by  
superposition of perturbative contributions,
we start with  the $\alpha$-representation 
for the relevant scalar diagram. The double distribution 
can be written similarly to Eq.(\ref{16uv}) 
\begin{equation}
M(u,v;s) = i m^2
\int_0^{\infty} \, 
\delta \left ( u - \frac{\alpha_4}{\lambda} \right   ) 
\delta \left ( v - \frac{\alpha_2}{\lambda}
\right   ) \, 
\exp \left \{  \frac{i }{\lambda} \left [ \alpha_2 \alpha_4 s 
+ \alpha_3 (\alpha_4  +\alpha_2 )m_{\pi}^2 \right ] -
i\lambda (m^2-i\epsilon)\right \} 
 \frac{d\alpha_2 d\alpha_3 d\alpha_4}{\lambda^2} \, .
\label{16suv} \end{equation}
Here, from the beginning  we take $p_1^2= p_2^2 = m_{\pi}^2$,
and add the overall factor $m^2$ to make the function dimensionless. 
Note, that for the triangle diagram we have $(1-u-v) = \alpha_3 /\lambda \equiv \beta_3$.
So, we write the 
scalar triangle version of the 2$\pi$DA as 
\begin{eqnarray}
\Phi(z,\zeta;s)  = i m^2    
\int_0^{\infty} d \lambda
 \, \int_0^1 d\beta_2 \int_0^1 d\beta_3 \, \theta (\beta_2+\beta_3 \leq 1)
\, \delta (z-\zeta \beta_3 - \beta_2) \nonumber \\
\exp \left \{  {i }{\lambda} \left [(1- \beta_2 -\beta_3) \beta_2 s 
+ \beta_3 (1-\beta_3  )m_{\pi}^2 - m^2+i\epsilon\right ] \right \} 
 \, .
\label{16Muv} \end{eqnarray}
Integrating over $\lambda$ and incorporating the $\delta$-function to calculate 
the $\beta_2$ integral, we get
\bea{Phizzetas}
\Phi(z,\zeta;s)  = 
   \int_0^{{\rm min}\{z/\zeta, \bar z/ \bar \zeta\}} d\beta_3 \, 
\left [1- (  z- \beta_3\zeta ) (\bar z  - \beta_3\bar \zeta) s/m^2 
- \beta_3 (1-\beta_3  )m_{\pi}^2 /m^2-i\epsilon\right ] ^{-1}
\ .
\end{eqnarray}
This representation explicitly  demonstrates the well-known 
fact (see, {\it e.g.,} \cite{Polyakov:1998td,Kivel:2002ia})  
   that $\Phi(z,\zeta;s)$ 
is non-analytic at the point $z=\zeta$. 
The integral can be taken in general case,
but it is  instructive to analyze  the simplest limit  $s=0\, , \, m_{\pi}^2=0$.
In this case, the result is the function 
\bea{Phzz00}
\Phi(z,\zeta;s=0)_{m_{\pi}^2=0}  = \frac{z}{\zeta} \, \theta (z<\zeta)
+ \frac{ 1- z}{ 1- \zeta}\,\theta (z>\zeta) 
\end{eqnarray}
that coincides  with a part of the pion DA  evolution
kernel. Its eigenfunctions
are the Gegenbauer polynomials $C_n^{3/2}(2\zeta -1)$ and   
the eigenvalues are $1/(n+1)(n+2)$ \cite{Lepage:1980fj,Efremov:1982dk}.
Hence, we can write
 \bea{Vzz}
 \Phi(z,\zeta;s=0)_{m_{\pi}^2=0} = 4 z(1-z) \sum_{n=0}^{\infty} 
 \frac{2n+3}{(n+1)^2(n+2)^2} \, C_n^{3/2}(2z -1) C_n^{3/2}(2\zeta -1)
 \ .
\end{eqnarray}

It  is convenient to write   2$\pi$DA as a sum over 
$z(1-z)C_n^{3/2}(2z -1)$, since these are the eigenfunctions of the 
evolution kernel. 
On the other hand, the combination $(2\zeta -1)$ is related to the
cosine of the angle between the pions' momenta,
so it is natural to expand the $\zeta$-dependence of $ \Phi(z,\zeta;s)$
in the Legendre polynomials $P_l (2\zeta -1)$.
Using the formula $$C_l^{3/2}(x) -  C_{l-2}^{3/2}(x)=(2l+1)P_l(x)\  ,$$
we  can write    $C_n^{3/2}(2\zeta -1)$ as a sum  of $P_l (2\zeta -1)$ and 
obtain
\bea{Vzznl}
 \Phi(z,\zeta;s=0)_{m_{\pi}^2=0} = 4 z(1-z) \sum_{n=0}^{\infty} 
 \frac{2n+3}{(n+1)^2(n+2)^2} \, C_n^{3/2}(2z -1) \sum_{l=0}^{n}
\, (2l+1) P_l (2\zeta -1)\, \frac{1+(-1)^{n-l}}{2} 
 \ .
\end{eqnarray}
This  expansion  has the structure of the general representation  
for 2$\pi$DAs proposed 
by Polyakov \cite{Polyakov:1998ze}. 
In specific  models,   only the first  
terms of the expansion are included. It is easy to check that, in our case, 
the exact result (\ref{Phzz00}) 
is well approximated by a few first terms of the Gegenbauer 
expansion (\ref{Vzz}), even in the vicinity of the non-analyticity 
point $z=\zeta$.  Polyakov \cite{Polyakov:1998ze} 
considers also a more complicated $s\neq 0$ case,
using the $\pi \pi$ scattering information 
to model  the $s$-dependence.
Recently, Kivel and Polyakov \cite{Kivel:2002ia} used chiral perturbation theory 
to include $O(m_{\pi}^2)$ corrections to the chiral limit.

Now we want  to show how one can 
incorporate information about usual (forward)
parton densities to build  models for 2$\pi$DAs.
To this end, it is convenient to 
 write the pion momenta as $p_1=P/2+r$ and $p_2 = P/2-r$
 (the plus-components are implied, but we  omit the $+$ 
 superscript here and below).
The quark momenta can be written then as 
\bea{k1k2}
k_1 = \frac{1+\alpha}{2} P + xr \  \  \  ,  \  \  \  k_2 = \frac{1-\alpha}{2} P - xr \ ,
\end{eqnarray}
where the variables $\alpha$ and $x$ are related to $u,v$ by $x=1-u-v$ and 
$\alpha = v-u$. The support region is $|\alpha| \leq 1-x$.
The 2$\pi$DA $\Phi (z,\zeta;s)$ is related to  the double distribution $F (x,\alpha;s)$ by 
 \bea{PhiF}
 \Phi(z,\zeta;s)= \int_0^1 dx \int_{-1+x}^{1-x}  
 d\alpha  \,  \delta (z-1/2-x(\zeta -1/2) -\alpha/2)\, 
 F (x,\alpha;s) \  .
\end{eqnarray} 
In this description, the total pair momentum $P$ is shared 
by the quarks in fractions $(1+\alpha)/2$ and $(1-\alpha)/2$,
while the relative momentum $r$ is carried by active quarks 
in fractions $x$ and $-x$. 
Hence, the relevant double distribution $F(x,\alpha;s)$ 
is the timelike analog of the function $f(x,\alpha;t)$  
considered in Section II.
In the forward limit,  it reduces to the usual parton densities:
\bea{red}
 \int_{-1+x}^{1-x}  F (x,\alpha;s=0) \, d\alpha = f (x) \  .
\end{eqnarray} 
In the $m_{\pi}^2=0$ case, we have $ F (x,\alpha;s=0) =\frac12 \theta (|\alpha|  \leq 1-x)$,
hence the integral in (\ref{red}) gives $(1-x)$, which is exactly the forward 
distribution for the scalar massless triangle. 
To  get a more realistic $f(x)$, we can use the $R(a)$-model described in Appendix B.
It gives 
\bea{PhiDD}
   \Phi^R(z,\zeta;s)= 
   \int_0^{\min\{z/\zeta, \bar z / \bar \zeta\}} dx \, 
   \int_0^{\infty} \exp \left\{-\rho \left [1- (z-x\zeta)(\bar z -x \bar \zeta)\, \frac{s}{m^2}
   -x\bar x \, \frac{m_{\pi}^2}{m^2} \right  ] \right \} 
   \, R\left (\rho\, \sqrt{(z-x\zeta)(\bar z -x \bar \zeta)}\,  \right ) 
   \, d\rho\, \, .
\end{eqnarray} 

According to  Appendix  B, the $R(a)$-model is equivalent  
to a sum of ``wave function overlap''  contributions of  Eq.(\ref{Bgen}) 
type, similar to those 
obtained within  the light-cone approaches (see, {\it e.g.,} \cite{Brodsky:2000xy,Diehl:2002he,Tiburzi:2002mn}).
However, the $R(a)$ construction has the advantage 
that it provides also a model for  2$\pi$DA.
In the standard light-cone formalisms,  the 2$\pi$DA 
  would involve  the $q \to \pi q$ vertices
  which cannot be interpreted as light-cone wave functions.

In the scalar triangle model,  the 2$\pi$DA $\Phi (z,\zeta;s=0)$ is obtained from the same 
DD $F (x,\alpha;s=0)=f (x,\alpha;t=0) $ which produces the $t=0$ OFPD 
\mbox{$H(\tilde x,\xi;t=0)$}.  Comparing  Eqs.(\ref{PhiF}) and  (\ref{OFPD}),
 we can formally write
 \bea{PhiH}
\Phi(z,\zeta;s=0) = H \left (\frac{2z-1}{2\zeta -1}\, , \, \frac1{2\zeta -1}\, ; \, t=0 \right ). 
 \end{eqnarray}
 The relation is even simpler 
 \bea{PhiHs}
\phi(\tilde x,\tilde \xi;s=0) = H \left ({\tilde x}/{\tilde \xi}\, , \, 
1/{ \tilde \xi}\, ; \, t=0 \right )  
 \end{eqnarray}
 for the  2$\pi$DA $\phi (\tilde x,\tilde \xi;s=0)$ 
 written in the symmetric variables $\tilde x =2z-1$ and $\tilde \xi = 2\zeta -1$.
   Since $|\tilde \xi| \leq 1$, the OFPD  $H ({\tilde x}/{\tilde \xi}\, , \, 
1/{ \tilde \xi}\, ; \, t=0  )$ is taken at the skewedness values 
with absolute magnitude larger than 1. 
 Hence, as suggested by Teryaev \cite{Teryaev:2001qm}, the 2$\pi$DA 
 may be treated as a continuation of OFPD into the $|\xi|>1$ 
 region. More precisely, $H|_{t=0}$ and $\Phi|_{s=0}$ may be treated 
 as $|\xi|<1$ and $|\xi|>1$ components of the same function.
 
 To make  parallel with $I=0$ and $I=1$ components of 2$\pi$DAs in QCD,
one should take  combinations 
\bea{Phisym}
\Phi^{\pm} (z,\zeta;s) = \frac12 \biggl [\Phi (z,\zeta;s) \pm \Phi (1-z,\zeta;s) \biggr ] \ ,
\end{eqnarray}
which are symmetric or antisymmetric with respect to the middle point $z=1/2$.
They are given by summation over even or odd $n$ in Eq.(\ref{Vzznl}).
Taking $s=0, m_{\pi}^2=0$ and fixing $R(a)$ in the same way as in Appendix B,
we obtained the curves shown in Fig.\ref{2-pion} .

\begin{figure}[t]
\begin{center}
\mbox{
   \epsfxsize=6cm
 \epsfysize=4cm
  \epsffile {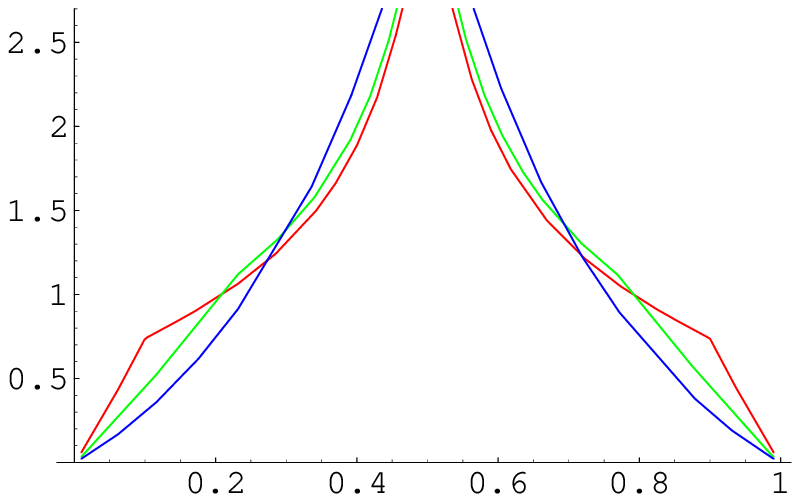} } \hspace{0cm}
\mbox{\epsfxsize=6cm
 \epsfysize=4cm
  \epsffile {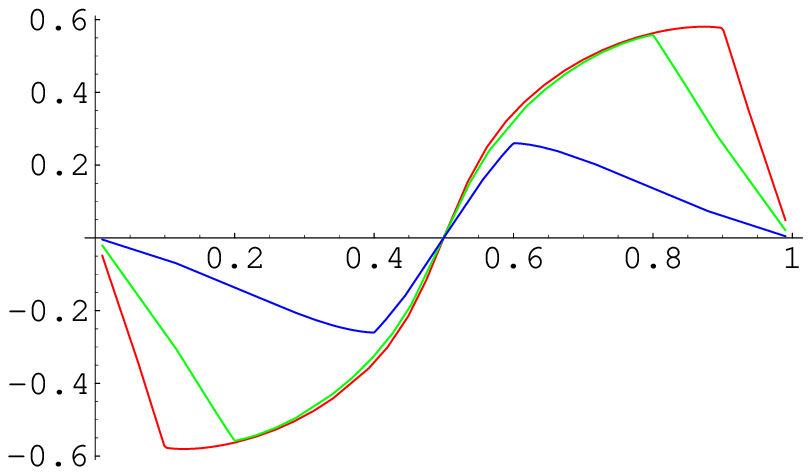} }
  \end{center}
  \vspace{3mm}
\caption{\label{2-pion}
Two-pion distribution amplitudes  $\Phi^{+} (z,\zeta;s=0)$  and $\Phi^{-} (z,\zeta;s=0)$
with $\zeta=0.1$, $0.2$, $0.4$ obtained in the scalar $R(a)$-model.
 The forward distribution was modeled by  $f^{R}(x)=\frac34 (1-x)/\sqrt{x}$.
}
\end{figure}

At $z=1/2$,
the symmetric function in this model is infinite. This result  is similar 
to singularity  of SPDs ${\cal F}_{\zeta} (X)$ for $X=\zeta/2$ observed 
in Appendix B.  It  reflects the fact that 
the profile of  $F(x,\alpha,s=0)$ in the $R(a)$-model 
becomes infinitely narrow as $x\to 0$.
Indeed,   for DDs   with infinitely narrow profile
for all $x$, {\it i.e.,}  for $F(x,\alpha,s=0) =f(x) \delta(\alpha)$, 
we would have $\Phi(z,\zeta;s=0) = f[(z-1/2)/(\zeta -1/2)]/(1-\zeta/2)$,
which gives infinite result for $z=1/2$ if $f(0) \to \infty$ .
The curves also have cusps for $z=\zeta$ and $z=1-\zeta$.
They appear because the DD $F(x,\alpha,s=0)$ of the $R(a)$-model
does not vanish at the  upper corner $x=0, \alpha=1$ of the support region.
This is because the  profile function for the lowest term 
of the $R(a)$ expansion is $h^{(0)}(x,y)=1/(1-x)$: unlike 
profile functions $h^{(n)}(x,y)$ with $n>0$, it does not vanish at the 
border lines $x+|\alpha|=1$.

Note that 2$\pi$DAs of the purely scalar model 
are symmetric with respect to the change 
$\{z\to 1-z, \zeta \to 1 - \zeta \}$ while 
in QCD the   2$\pi$DAs  describing  transition of  spin-1/2 
quarks into pions changes sign after this transformation. 
The triangle perturbative contributions for this  case
were considered a few years ago 
by Polyakov and Weiss \cite{Polyakov:1998td}.
We plan to extend  their 
calculation by combining it with the ideas 
of  the present  paper.

\end{appendix}



\begin{thebibliography}{99}



\bibitem{Muller:1998fv}
D.~M{\"u}ller, D.~Robaschik, B.~Geyer, F.~M.~Dittes and J.~Ho{\v r}ej{\v s}{\'\i},
Fortsch.\ Phys.\  {\bf 42}, 101 (1994)
[arXiv:hep-ph/9812448].

\bibitem{Ji:1996ek}
X.~D.~Ji,
Phys.\ Rev.\ Lett.\  {\bf 78}, 610 (1997)
[arXiv:hep-ph/9603249].

\bibitem{Radyushkin:1996nd}
A.~V.~Radyushkin,
Phys.\ Lett.\ B {\bf 380}, 417 (1996)
[arXiv:hep-ph/9604317],
Phys.\ Lett.\ B {\bf 385}, 333 (1996)
[arXiv:hep-ph/9605431].

\bibitem{Ji:1996nm}
X.~D.~Ji,
Phys.\ Rev.\ D {\bf 55}, 7114 (1997)
[arXiv:hep-ph/9609381].
 


\bibitem{Radyushkin:1997ki}
A.~V.~Radyushkin,
Phys.\ Rev.\ D {\bf 56}, 5524 (1997)
[arXiv:hep-ph/9704207].


\bibitem{Goeke:2001tz}
K.~Goeke, M.~V.~Polyakov and M.~Vanderhaeghen,
Prog.\ Part.\ Nucl.\ Phys.\  {\bf 47}, 401 (2001)
[arXiv:hep-ph/0106012].


\bibitem{Radyushkin:1998rt}
A.~V.~Radyushkin,
Phys.\ Rev.\ D {\bf 58}, 114008 (1998)
[arXiv:hep-ph/9803316].


\bibitem{Diehl:1998kh}
M.~Diehl, T.~Feldmann, R.~Jakob and P.~Kroll,
Eur.\ Phys.\ J.\ C {\bf 8}, 409 (1999)
[arXiv:hep-ph/9811253].

\bibitem{Radyushkin:1998es}
A.~V.~Radyushkin,
Phys.\ Rev.\ D {\bf 59}, 014030 (1999)
[arXiv:hep-ph/9805342].

\bibitem{huang} G. P. Lepage, S. J. Brodsky, T. Huang and P. B. Mackenzie,
In: {\it Particles and Fields 2}, Eds. A. Z. Capri and A. N. Kamal, Plenum  
Press, New York, 1981.
 


\bibitem{Frederico:2001qy}
T.~Frederico and H.~C.~Pauli,
Phys.\ Rev.\ D {\bf 64}, 054007 (2001)
[arXiv:hep-ph/0103233].

\bibitem{Pauli:2001uf}
H.~C.~Pauli and A.~Mukherjee,
Int.\ J.\ Mod.\ Phys.\ A {\bf 16}, 4351 (2001)
[arXiv:hep-ph/0103150],
H.~C.~Pauli,
arXiv:hep-ph/0107302.
 

\bibitem{schlumpf} F. Schlumpf, Phys. Rev. {\bf D47}, 4114 (1993); {\bf D48},
4478 (1993).


\bibitem{Ji:1998pc}
X.~D.~Ji,
J.\ Phys.\ G {\bf 24}, 1181 (1998)
[arXiv:hep-ph/9807358].

\bibitem{Polyakov:1999gs}
M.~V.~Polyakov and C.~Weiss,
Phys.\ Rev.\ D {\bf 60}, 114017 (1999)
[arXiv:hep-ph/9902451].

\bibitem{Mankiewicz:1997uy}
L.~Mankiewicz, G.~Piller and T.~Weigl,
Eur.\ Phys.\ J.\ C {\bf 5}, 119 (1998)
[arXiv:hep-ph/9711227].

\bibitem{Drell:1969km}
S.~D.~Drell and T.~M.~Yan,
Phys.\ Rev.\ Lett.\  {\bf 24}, 181 (1970).


\bibitem{Barone:ej}
V.~Barone, M.~Genovese, N.~N.~Nikolaev, E.~Predazzi and B.~G.~Zakharov,
Z.\ Phys.\ C {\bf 58}, 541 (1993).



\bibitem{Terentev:jk}
M.~V.~Terentev,
Sov.\ J.\ Nucl.\ Phys.\  {\bf 24}, 106 (1976)
[Yad.\ Fiz.\  {\bf 24}, 207 (1976)].

\bibitem{Amendolia:1984nz}
S.~R.~Amendolia {\it et al.},
Phys.\ Lett.\ B {\bf 146}, 116 (1984).




\bibitem{Volmer:2000ek}
J.~Volmer {\it et al.}  [The Jefferson Lab F(pi) Collaboration],
Phys.\ Rev.\ Lett.\  {\bf 86}, 1713 (2001)
[arXiv:nucl-ex/0010009].




\bibitem{Brodsky:1989pv}
S.~J.~Brodsky and G.~P.~Lepage,
 in: A. H. Mueller (Ed.),
Perturbative Quantum Chromodynamics, World Scientific, 1989.



\bibitem{Polyakov:1998ze}
M.~V.~Polyakov,
Nucl.\ Phys.\ B {\bf 555}, 231 (1999)
[arXiv:hep-ph/9809483].



\bibitem{Tiburzi:2002sxyz}
B.~C.~Tiburzi and G.~A.~Miller,
arXiv:hep-ph/0212238 .



\bibitem{Pobylitsa:2002vw}
P.~V.~Pobylitsa,
arXiv:hep-ph/0210238.




\bibitem{Martin:1997wy}
A.~D.~Martin and M.~G.~Ryskin,
Phys.\ Rev.\ D {\bf 57}, 6692 (1998)
[arXiv:hep-ph/9711371].




\bibitem{Pire:1998nw}
B.~Pire, J.~Soffer and O.~Teryaev,
Eur.\ Phys.\ J.\ C {\bf 8}, 103 (1999)
[arXiv:hep-ph/9804284].

\bibitem{Diehl:2000xz}
M.~Diehl, T.~Feldmann, R.~Jakob and P.~Kroll,
Nucl.\ Phys.\ B {\bf 596}, 33 (2001)
[Erratum-ibid.\ B {\bf 605}, 647 (2001)]
[arXiv:hep-ph/0009255].

\bibitem{Pobylitsa:2001nt}
P.~V.~Pobylitsa,
Phys.\ Rev.\ D {\bf 65}, 077504 (2002)
[arXiv:hep-ph/0112322].


\bibitem{Pobylitsa:2002gw}
P.~V.~Pobylitsa,
Phys.\ Rev.\ D {\bf 65}, 114015 (2002)
[arXiv:hep-ph/0201030].



\bibitem{Diehl:2002he}
M.~Diehl,
Eur.\ Phys.\ J.\ C {\bf 25}, 223 (2002)
[arXiv:hep-ph/0205208].




%
\bibitem{Pobylitsa:2002vi}
P.~V.~Pobylitsa,
arXiv:hep-ph/0210150.


\bibitem{Diehl:1998dk}
M.~Diehl, T.~Gousset, B.~Pire and O.~Teryaev,
Phys.\ Rev.\ Lett.\  {\bf 81}, 1782 (1998)
[arXiv:hep-ph/9805380].


\bibitem{Diehl:2001fv}
M.~Diehl, P.~Kroll and C.~Vogt,
Phys.\ Lett.\ B {\bf 532}, 99 (2002)
[arXiv:hep-ph/0112274].


\bibitem{Teryaev:2001qm}
O.~V.~Teryaev,
Phys.\ Lett.\ B {\bf 510}, 125 (2001)
[arXiv:hep-ph/0102303].

\bibitem{Kivel:2002ia}
N.~Kivel and M.~V.~Polyakov,
arXiv:hep-ph/0203264.

\bibitem{Polyakov:1998td}
M.~V.~Polyakov and C.~Weiss,
Phys.\ Rev.\ D {\bf 59}, 091502 (1999)
[arXiv:hep-ph/9806390].



\bibitem{Lepage:1980fj}
G.~P.~Lepage and S.~J.~Brodsky,
Phys.\ Rev.\ D {\bf 22}, 2157 (1980).

\bibitem{Efremov:1982dk}
A.~V.~Efremov, V.~A.~Nesterenko and A.~V.~Radyushkin,
Nuovo Cim.\ A {\bf 76}, 122 (1983).


\bibitem{Brodsky:2000xy}
S.~J.~Brodsky, M.~Diehl and D.~S.~Hwang,
Nucl.\ Phys.\ B {\bf 596}, 99 (2001)
[arXiv:hep-ph/0009254].

\bibitem{Tiburzi:2002mn}
B.~C.~Tiburzi and G.~A.~Miller,
arXiv:hep-ph/0205109.


\end{thebibliography}
\end{document}